\documentclass[usedcolumn,usenatbib]{mn2e}
\usepackage{psfig}

\title[The Millennium Galaxy Catalogue]{The Millennium Galaxy
Catalogue: ${\bf 16 \le B_{\rm MGC} < 24}$ galaxy counts and the
calibration of the local galaxy luminosity function}

\author[J.~Liske et al.]{J.~Liske$^{1}$\thanks{E-mail:
jol@roe.ac.uk}, D.~J.~Lemon$^2$, S.~P.~Driver$^3$,
N.~J.~G.~Cross$^4$ and W.~J.~Couch$^5$\\
$^1$Institute for Astronomy, University of Edinburgh, Royal
Observatory, Blackford Hill, Edinburgh EH9 3HJ\\
$^2$School of Physics and Astronomy, University of St Andrews, North
Haugh, St Andrews KY16 9SS\\
$^3$Research School of Astronomy \& Astrophysics, Australian National
University, Cotter Road, Weston, ACT 2611, Australia\\
$^4$Department of Physics and Astronomy, Johns Hopkins University,
3400 North Charles Street, Baltimore, MD 21218-2686, USA\\
$^5$School of Physics, University of New South Wales, Sydney, NSW
2052, Australia}

\date{Accepted
...... Received .....}

\pagerange{\pageref{firstpage}--\pageref{lastpage}}

\pubyear{2003}

\newcommand{\mpass}{~mag~arcsec$^{-2}$}
\newcommand{\bmgc}{B_{\rm MGC}}
\newcommand{\bdmgc}{B^\prime_{\rm MGC}}
\newcommand{\bj}{b_{\rm J}}
\newcommand{\eref}[1]{(\ref{#1})}

\begin{document}

\label{firstpage}

\maketitle

\begin{abstract}
The Millennium Galaxy Catalogue (MGC) is a $37.5$~deg$^2$, medium-deep,
$B$-band imaging survey along the celestial equator, taken with the
Wide Field Camera on the Isaac Newton Telescope. The survey region is
contained within the regions of both the Two Degree Field
Galaxy Redshift Survey (2dFGRS) and the Sloan Digital Sky Survey Early
Data Release (SDSS-EDR). The survey has a uniform isophotal detection
limit of $26$\mpass\ and it provides a robust, well-defined catalogue
of stars and galaxies in the range $16 \le \bmgc < 24$~mag.

Here we describe the survey strategy, the photometric and astrometric
calibration, source detection and analysis, and present the galaxy
number counts that connect the bright and faint galaxy populations
within a single survey. We argue that these counts represent the state
of the art and use them to constrain the normalizations ($\phi^*$) of
a number of recent estimates of the local galaxy luminosity
function. We find that the 2dFGRS, SDSS Commissioning Data (CD), ESO
Slice Project, Century Survey, Durham/UKST, Mt Stromlo/APM, SSRS2, and
NOG luminosity functions require a revision of their published
$\phi^*$ values by factors of $1.05\pm0.05$, $0.76\pm0.10$,
$1.02\pm0.22$, $1.02\pm0.16$, $1.16\pm0.28$, $1.75\pm0.37$,
$1.40\pm0.26$ and $1.01\pm0.39$, respectively. After renormalizing the
galaxy luminosity functions we find a mean local $\bj$ luminosity
density of $\overline{j_{\bj}} = (1.986 \pm 0.031) \times 10^8 \; h \;
L_{\odot}$~Mpc$^{-3}$.\footnotemark
\end{abstract}


\footnotetext{We use $H_0 = 100 \; h$~km~s$^{-1}$~Mpc$^{-1}$
throughout this paper.}

\section{Introduction} \label{introduction}
Our understanding of the local universe and the local galaxy
population originates primarily from the all-sky photographic Schmidt
surveys and the established catalogues of bright galaxies derived from
them, such as the Catalogue of Galaxies and Clusters of Galaxies
\citep{Zwicky68}, the Morphological Catalogue of Galaxies
\citep{Vorontsov74}, the Uppsala General Catalogue of Galaxies
\citep{Nilson73}, the ESO/Uppsala Catalogue
\citep{Lauberts82,Lauberts89}, the Southern Galaxy Catalogue
\citep*{Corwin85}, the Catalogue of Principal Galaxies
\citep{Paturel89}, the Edinburgh/Durham Southern Galaxy Catalogue
\citep*{Heydon89}, the APM catalogue \citep{Maddox90}, the Third
Reference Catalogue of Bright Galaxies \citep{deVaucouleurs91} and the
SuperCOSMOS Sky Survey \citep{Hambly01}. While these catalogues have
provided invaluable information and insight, uncertainty remains as to
their completeness, particularly for low surface brightness and
compact galaxies \citep{Disney76,Sprayberry97,Impey97,Drinkwater99}.
In addition there are concerns as to the photometric accuracy
\citep*[e.g.][]{Metcalfe95b}, the susceptibility to scale errors
\citep{Bertin97}, plate-to-plate variations \citep{Cross03} and
dynamic range.

These photographic-based catalogues have been the starting point for
numerous spectroscopic surveys aimed at measuring the local space
density of galaxies (i.e.\ the local galaxy luminosity function). The
space density of galaxies is our fundamental census of the local
contents of space and therefore a crucial constraint for models of
galaxy formation \citep[e.g.][]{White91,Cole00,Pearce01}. If the
imaging catalogues are in omission and/or photometrically inaccurate
then regardless of the completeness of the spectroscopic surveys our
insight into the galaxy population will be incomplete and most likely
biased against specific galaxy types.

Over the past two decades there have been numerous estimates of the
local galaxy luminosity function (e.g.\ EEP, \citealt*{Efstathiou88}; Mt
Stromlo/APM, \citealt{Loveday92}; Autofib, \citealt{Ellis96}; ESP,
\citealt{Zucca97}; SSRS2, \citealt{Marzke98}; Durham/UKST,
\citealt{Ratcliffe98}; SDSS-CD, \citealt{Blanton01}; 2dFGRS,
\citealt{Norberg02}) and of the three-parameter Schechter function used
to represent it \citep{Schechter76}. Typically the surveys agree
broadly on the faint end slope ($\alpha$, $\Delta \alpha \approx \pm
0.15$) but show a marked variation in the characteristic luminosity
($L^*$, $\Delta L^* \approx 40$ per cent) and normalization ($\phi^*$,
$\Delta \phi^* \approx 50$ per cent). The uncertainties in the
Schechter parameters result in an uncertainty of $> 60$ per cent in
the local luminosity density, $j = \phi^*  L^*  \Gamma(\alpha+2)$.

This uncertainty is usually expressed as the normalization problem
which has been somewhat overshadowed by the more notorious faint blue
galaxy problem \citep{Koo92,Ellis97}. The latter describes the
inability of basic galaxy number count models to predict the numbers
of galaxies seen at faint magnitudes ($22 < B < 28$~mag) in the deep
pencil beam CCD-based surveys
\citep[e.g.][]{Tyson88,Metcalfe95,Metcalfe01}. The lesser known
normalization problem describes the inability of number count models
to explain the galaxy counts even at bright magnitudes ($18 < B <
20$~mag) by as much as a factor of $2$ \citep*[see discussions
in][]{Shanks84,Driver95,Marzke98,Cohen03}. In many ways the
normalization problem is the more fundamental: while luminosity
evolution, cosmology and/or dwarf galaxies can be, and have been,
invoked in varying mixtures to explain the faint blue galaxy problem
\citep*[e.g.][]{Broadhurst88,Babul92,Phillipps95,Ferguson98}, none of
these can be used to resolve the normalization problem.

\begin{figure}
\psfig{file=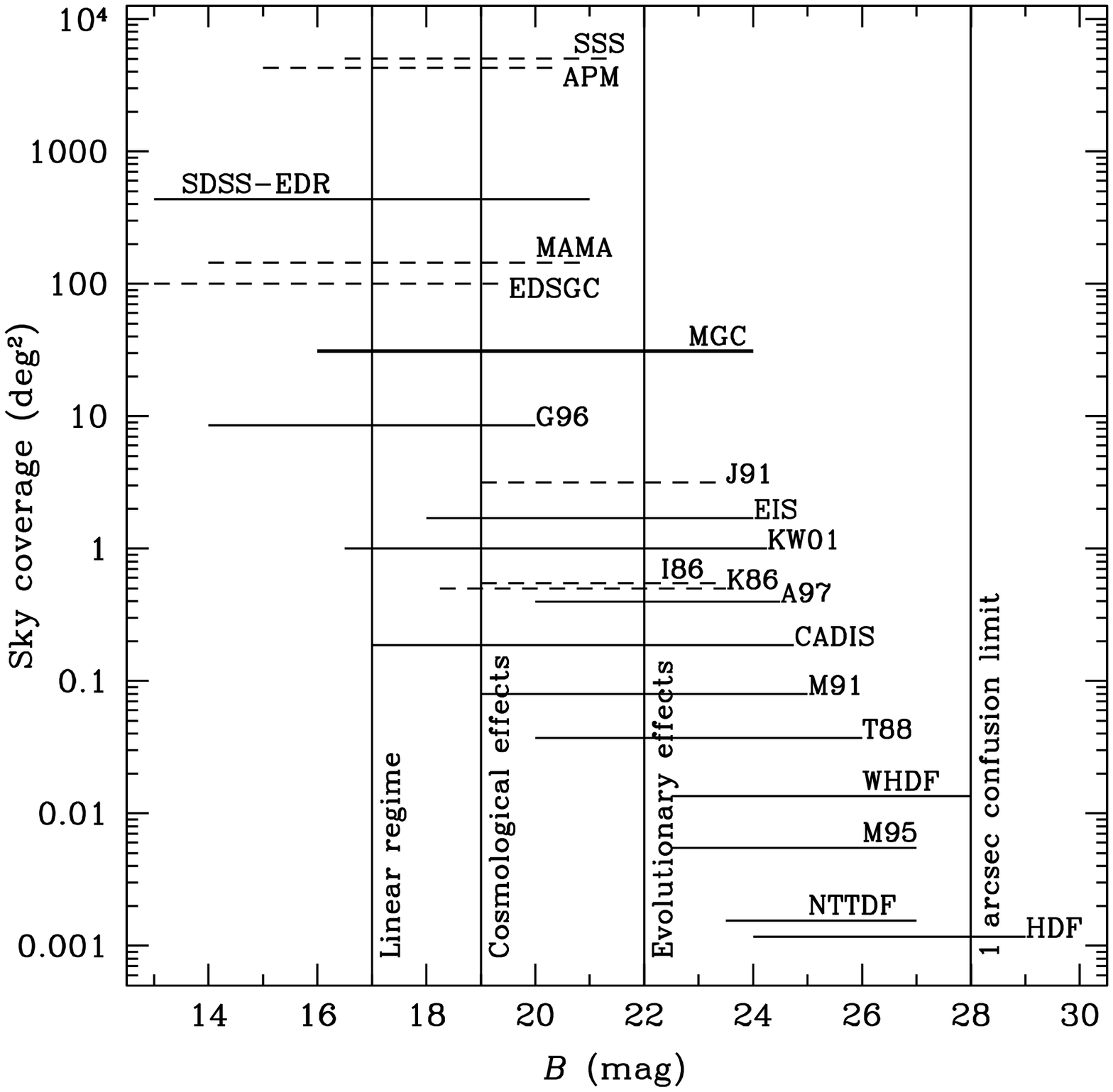,width=\columnwidth}
\caption{The magnitude ranges and survey areas spanned by some
previous number count publications. Surveys based on photographic
material are shown with a dashed line. The vertical lines show various
transition regions where various effects start to dominate the galaxy
counts. Key: SSS \citep{Hambly01}, APM \citep{Maddox90b}, SDSS-EDR
\citep{Yasuda01}, MAMA \citep{Bertin97}, EDSGC \citep{Heydon89}, MGC
(this work), G96 \citep{Gardner96}, J91 \citep{Jones91}, EIS
\citep{Prandoni99}, KW01 \citep{Kummel01}, I86 \citep*{Infante86}, K86
\citep{Koo86}, A97 \citep{Arnouts97}, CADIS \citep{Huang01}, M91
\citep{Metcalfe91}, T88 \citep{Tyson88}, WHDF \citep{Metcalfe01}, M95
\citep{Metcalfe95}, NTTDF \citep{Arnouts99}, HDF \citep{Williams96}.}
\label{surveys}
\end{figure}

In the past the problem was typically circumvented by renormalizing
the number count models to the range $18 < B < 20$~mag
\citep*[e.g.][]{Driver94,Metcalfe95,Driver95,Driver98,Marzke98,Metcalfe01}.
The justification was that the bright galaxy catalogues, on which the
luminosity function measurements are based, are shallow and therefore
susceptible to local clustering. However the crucial normalization
range typically occurs at the faint limit of the photographic surveys
(where the photometry and completeness are more likely to be a
problem) and at the bright end of the pencil beam CCD surveys (where
statistics are poor). While convenient, the clustering explanation
overlooks two more worrisome possibilities: gross photometric errors
and/or gross incompleteness in the local catalogues. If either of
these two latter explanations play a part this will have important
consequences for the new-generation spectroscopic surveys, namely the
2dFGRS \citep{Colless01} and the SDSS \citep{York00}. The input
catalogue of the 2dFGRS is an extensively revised version of the
photographic APM survey (which is known to show a peculiar steepening
in its galaxy counts at bright magnitudes, \citealt{Maddox90b}), with
zero-point and scale-error corrections from a variety of sources
including the 2MASS $K$-band survey and the data presented in this
paper (see \citealp{Norberg02} for details). In the case of the SDSS --
which leaps forward in terms of dynamic range, uniformity and
wavelength coverage -- the effective exposure time is relatively short
($54$~s) and the isophotal detection limit is comparable to that of
the photographic surveys. Hence while issues of photometric accuracy
should be resolved the question mark of completeness may remain.

\begin{figure}
\centerline{\psfig{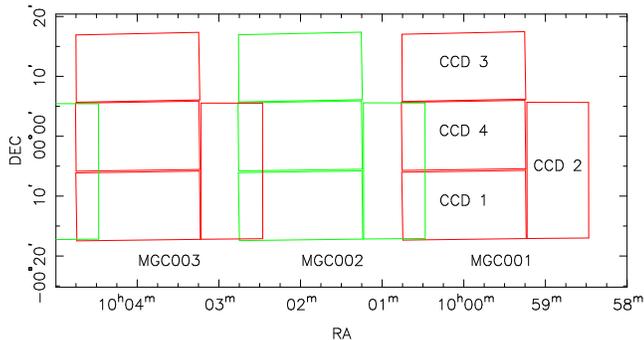}}
\caption{Outline of MGC fields 1, 2 and 3 and the arrangement of the
CCDs.}
\label{outline}
\end{figure}

To address the above problems within a single, well-defined dataset we
require a survey that is reasonably deep and yet has a large enough
solid angle to provide accurate and statistically significant galaxy
counts over the crucial normalization range. Furthermore, the survey's
photometry must be accurate and its completeness high, i.e.\ it must
probe to low surface brightnesses. Fig.~\ref{surveys} shows a number
of imaging surveys in terms of their sky coverage and magnitude
range. Dashed and solid lines indicate photographic and CCD-based
surveys, respectively. Typically the faint surveys are CCD-based while
the local surveys are photographic (with the notable and recent
exception of the SDSS-EDR, \citealt{Stoughton02}). While the CCD
surveys make significant improvements in surface brightness and
magnitude limits their sky coverage is small. It is only very recently
that large CCD mosaics such as the Wide Field Camera
\citep[WFC,][]{Irwin01} and the SDSS instrument \citep{Gunn98} have
been constructed that now allow a large area of sky to be surveyed
within a realistic time frame.

In this paper we present the Millennium Galaxy Catalogue (MGC,
Sections \ref{data}--\ref{sellimits}). The MGC represents a new
medium-deep, wide-angle galaxy resource, which firmly connects the
local and distant universe within a single dataset 
(cf.\ Fig~\ref{surveys}). In Section \ref{counts} we produce the galaxy
number counts spanning the range $16 \le \bmgc < 24$~mag. We then
focus on the normalization problem by comparing our counts over the
range $16 \le \bmgc < 20$~mag to the predictions of a number of local
luminosity function estimates in Section \ref{lumfs}. Our counts
provide stringent constraints on the normalization of the luminosity
function and hence on the local luminosity density. Our conclusions are
given in Section \ref{conclusions}.

The 2dFGRS and SDSS will essentially supersede all previous redshift
surveys and therefore it is important to verify their photometric
accuracy and completeness on as large a scale as possible. We will
provide a detailed comparison of the 2dFGRS and SDSS-EDR imaging
catalogues with the MGC in a companion paper \citep{Cross03}.

\begin{figure}
\psfig{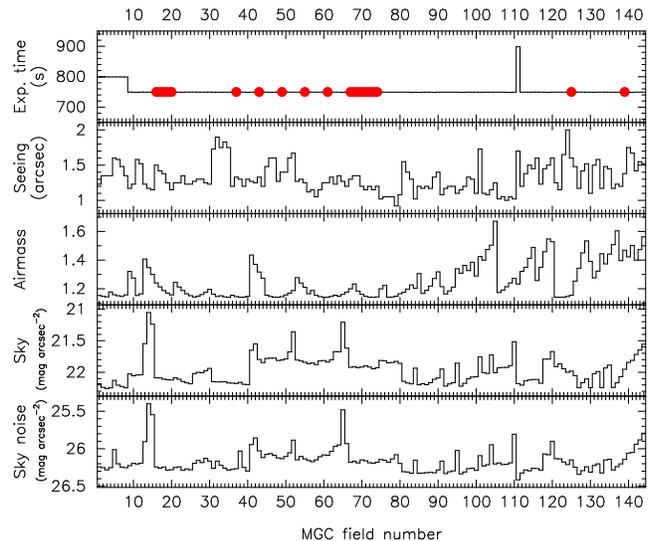}
\caption{Summary of the data quality across the MGC survey strip. The
dots in the uppermost panel indicate the location of the photometric
calibration fields. The sky and sky noise parameters were calculated
from the measured mode and rms of the background pixel value
distribution.}
\label{obsstat}
\end{figure}

A more long-term aim of the MGC project is to provide structural
information on the galaxy population around the crucial normalization
point ($16 < B < 20$~mag). It is ironic that since the advent of the
Hubble Space Telescope we have a greater understanding of the
morphological mix of galaxies at faint magnitudes than at bright
magnitudes. For example, \citet{Driver98} and \citet{Cohen03}
published morphological galaxy counts spanning the range $21 < B_{\rm
F450W} < 26$~mag, yet no reliable morphological galaxy counts at
brighter magnitudes exist. Consequently, no accurate local
morphological luminosity functions exist (compare, for example, the
conflicting results of \citealt{Loveday92} and \citealt{Marzke98}) and
the evolution of the different morphological types cannot be
accurately constrained. The MGC will enable us to remedy this
situation as it allows morphological classification and the extraction
of structural parameters to $\bmgc = 20$~mag.

The data and catalogues presented in this paper are publically
available at http://www.roe.ac.uk/$\sim$jol/mgc/.

\section{The data} \label{data}
\subsection{The Wide Field Camera}
All data frames where taken using the Wide Field Camera (WFC). The WFC
is mounted at prime focus on the $2.5\,$m Isaac Newton Telescope (INT)
situated at La Palma. The WFC is a mosaic of four 4k$\times$2k thinned
EEV CCDs with a smaller 2k$\times$2k Loral CCD which is used for
auto-guiding. Each of the science CCDs measures $2048 \times 4100$
pixels with a pixel scale of $0.333$~arcsec/pixel -- this gives a
total sky coverage of $0.287$~deg$^2$ per pointing. The four science
chips are arranged as shown in Fig.~\ref{outline}. Full details of the
WFC are provided at
http://www.ast.cam.ac.uk/$\sim$wfcsur/technical.html \citep[see
also][]{Irwin01}.

\begin{figure*}
\centerline{\psfig{file=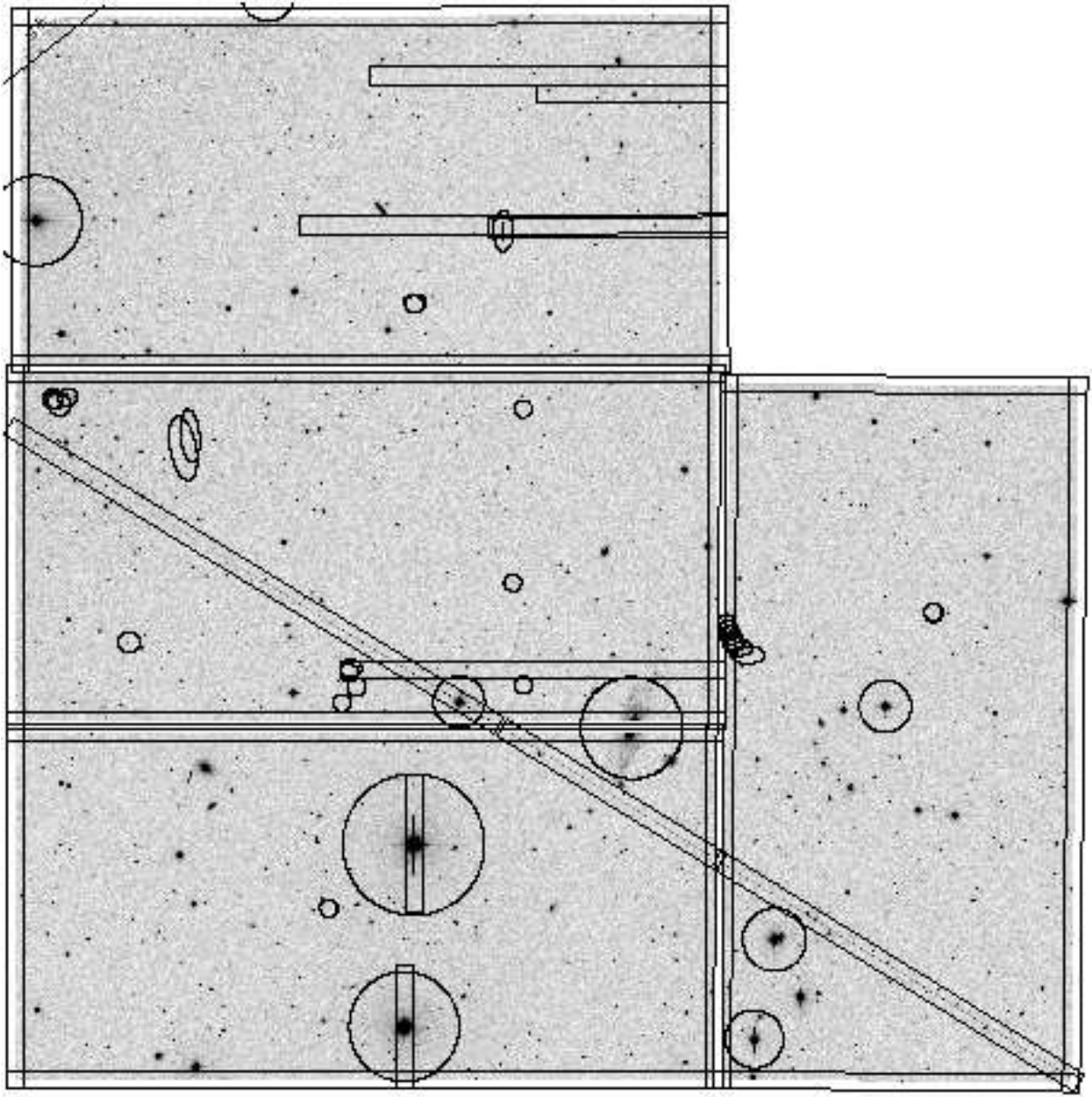,width=0.9\textwidth}}
\caption{The data of MGC field 36. Overlaid are the exclusion regions
due to CCD edges, the vignetted corner of CCD 3 (top), CCD defects,
bright stars and galaxies, satellite trails and diffraction spikes
(see Section \ref{exregions}). The exclusion regions reduce the
effective area covered by this pointing from $0.260$~deg$^2$ to
$0.217$~deg$^2$.}
\label{mgc036}
\end{figure*}

\subsection{The observations}
The data constituting the MGC comprise $144$ overlapping fields
forming a $35$~arcmin wide equatorial strip from $9^{\rm h} 58^{\rm m}
28^{\rm s}$ to $14^{\rm h} 46^{\rm m} 45^{\rm s}$ (J2000). The
observations were taken during $4$ observing runs, 1999 March 15--16,
1999 April 16--17, 1999 June 6--13 and 2000 March 26--April 4. Each
field was observed for a single $750$~s exposure through a Kitt Peak
National Observatory $B$ filter. Field $1$ is centered on RA $=
10^{\rm h} 00^{\rm m} 00^{\rm s}$, Dec $= 00\degr 00\arcmin 00\arcsec$
(J2000) and field $144$ is centered on RA $= 14^{\rm h} 46^{\rm m}
00^{\rm s}$, Dec $= 00\degr 00\arcmin 00\arcsec$ (J2000). Hence each
field is offset from the previous by $30$~arcmin along the equatorial
great circle. Fig.~\ref{outline} shows the survey outline for the
first three pointings. Note the substantial overlap between
neighbouring fields. The survey region was chosen because it is
contained within both the 2dFGRS \citep{Colless01} and the SDSS-EDR
\citep{Stoughton02} regions, thus providing redshifts and colours for
the brighter galaxies and allowing a detailed check of the photometry
and completeness of these surveys.

All observations were taken during dark or grey time through variable
conditions. The seeing ranged from $0.9$ to $2.0$~arcsec with the
median seeing at $1.3$~arcsec. The air masses ranged from $1.141$ to
$1.672$. Fig.~\ref{obsstat} shows a summary of the general observing
conditions across the survey.

As much of the data were collected during clear but non-photometric
nights it was necessary to dedicate a single pristine photometric
night (2000 March 30) to obtaining suitable calibration data at
various stages along the survey strip. In total $20$ MGC fields were
observed during the photometric night. Of these, six were only $100$~s
exposures as they had already been observed previously. These
observations were interspersed with $10$~s observations of standard
stars spanning a wide range in airmass. The standard stars where
taken from the \citet{Landolt92} standard areas SA98, SA101, SA104 and
SA107.

Some science frames were later found to be of too poor a quality to be
useful and these were re-observed: the first eight fields were
replaced with two $400$~s exposures each and field $111$ is a single
$900$~s exposure.

\subsection{Data reduction and astrometry}
All the preliminary data reduction -- flat-fielding, bias correction
and astrometric calibration -- was done by the Cambridge Astronomy
Survey Unit (CASU) and full details of this process are provided by
\citet{Irwin01}. Briefly, a number of bias frames are collected each
night and the median is subtracted from the data. All data (including
flat-fields) are corrected for a known non-linearity. A twilight
flat-field is taken during evening and morning twilight (when
possible) and a median flat-field derived for that particular run is
divided into each data frame. After this process an initial
astrometric calibration is made to the HST Guide Star Catalogue.
Finally the frames are matched to the APM catalogue which itself is
calibrated onto the Tycho-2 astrometric system. Fig.~\ref{mgc036}
shows the final reduced image for one of our pointings, field 36. It
illustrates problems with satellite trails, CCD defects, gaps between
CCDs, etc.

\begin{figure}
\psfig{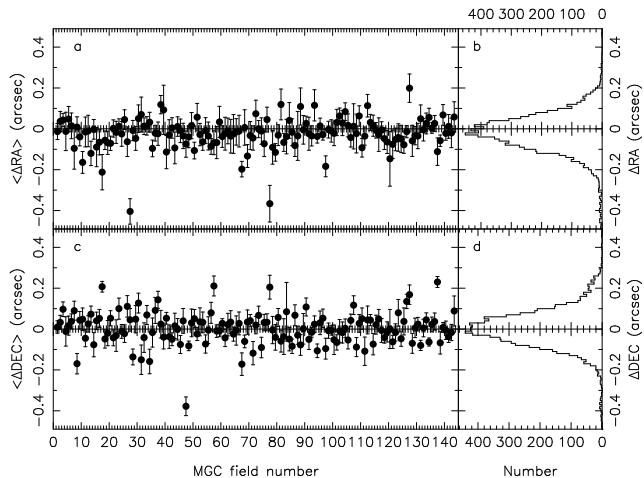}
\caption{(a) Median RA differences of doubly detected objects with $17
\le \bdmgc \le 21$~mag in overlap regions. The error bars are 68 per
cent ranges. (b) Frequency distribution of individual $\Delta$RA
values. This distribution has a mean and rms of $-0.016$ and
$0.083$~arcsec, respectively. (c) Same as (a) for Dec differences. (d)
Same as (b) for $\Delta$Dec. This distribution has a mean and rms of
$0.006$ and $0.084$~arcsec.}
\label{radecoff}
\end{figure}

To assess the final astrometric accuracy we compared the positions of
doubly detected objects in regions where neighbouring fields overlap
(cf.\ Fig.~\ref{outline}). The overlap regions are of size $\sim
0.027$~deg$^2$ and each contains $\sim 60$ objects in the range $17
\le \bdmgc \le 21$~mag,\footnote{Here and in Section \ref{calibration},
$\bdmgc$ refers to the KRON magnitude, uncorrected for Galactic
extinction (cf.\ Section \ref{photometry}).} where detections can be
easily and confidently matched. Fig.~\ref{radecoff} shows the median
positional differences for these objects for each overlap region as
well as the overall $\Delta$RA and $\Delta$Dec distributions. We find
that both of these distributions have an rms of $\pm
0.08$~arcsec. Note, however, that they are slightly but significantly
offset from zero which is most likely due to residual radial
distortions.

\section{Photometric calibration} \label{calibration}
\begin{figure}
\psfig{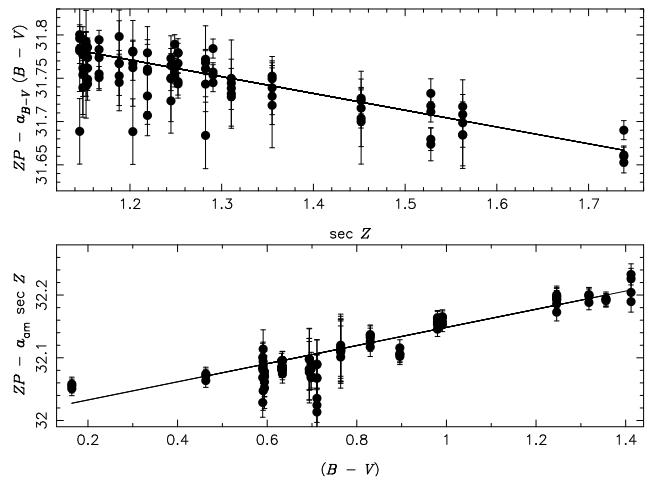}
\caption{Upper panel: colour-corrected zero-points versus
airmass. Lower panel: airmass-corrected zero-points versus
colour. The solid lines show the best fit 
(cf.\ equation~\ref{zpfit}). The error bars include contributions from
photon counting, sky subtraction, read-out noise and the uncertainty
in $B$. A further systematic error of $0.01$~mag (added in quadrature)
was needed to achieve an acceptable fit.}
\label{stdzp}
\end{figure}

As mentioned previously, four \citet{Landolt92} standard star fields
were observed at a range of air masses throughout the course of the
photometric night. For each observation of each standard star we
computed a zero-point
\begin{equation}
ZP_{\rm std} = B + 2.5 \log f,
\end{equation}
where $B$ was taken from \citet{Landolt92} and $f$ is the flux of the
star as measured from the data. We then fitted these zero-points with
a double linear function in airmass ($= \sec Z$, where $Z$ is the
zenith distance), and colour,
\begin{equation}
\label{zpfit}
ZP_{\rm std} = a + a_{\rm am} \: \sec Z + a_{B-V} (B-V),
\end{equation}
where $(B-V)$ is again taken from \citet{Landolt92}. In
Fig.~\ref{stdzp} we show the data and the fit. The residuals have an
rms of $0.02$~mag and show no obvious trend with airmass, colour, $B$
or time of observation. Note, however, that a systematic error of
$0.01$~mag was needed to achieve an acceptable fit. From the lower
panel of Fig.~\ref{stdzp} we can see that for several stars multiple
observations of the same star give consistently high or low results.
This may indicate that these stars are slightly variable or that the
errors on the \citet{Landolt92} photometry have been underestimated.

\begin{figure}
\psfig{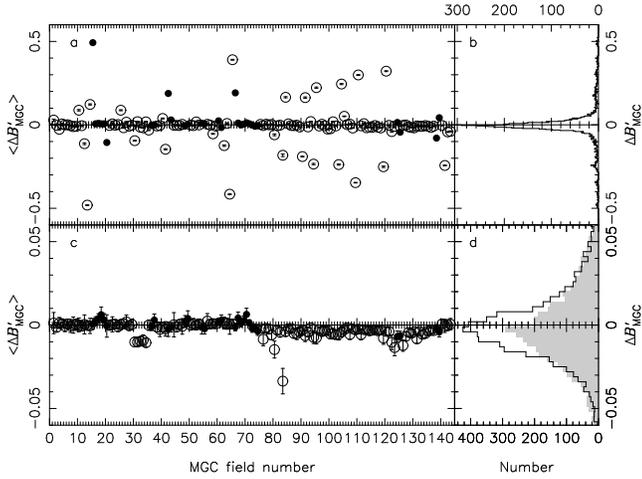}
\caption{(a) Median magnitude differences of doubly detected objects
with $17 \le \bdmgc \le 21$~mag and stellaricity $> 0.5$ in overlap
regions, using initial zero-points. Solid points indicate overlap
regions involving at least one photometric observation. The error bars
are $68$ per cent ranges divided by the square root of the number of
objects (on average $35$). (b) Frequency distribution of all of the
individual $\Delta\bdmgc$ values. (c) Same as (a) using the final
zero-points. Note that the scale of the $\langle \Delta\bdmgc
\rangle$-axis is expanded by a factor of $10$ compared with (a). (d)
The solid line shows the frequency distribution of the individual
$\Delta\bdmgc$ values using the final zero-points. This distribution
has a mean of $-0.002$~mag and an rms of $0.023$~mag. For comparison,
the shaded histogram is the same as that shown in (b).}
\label{dm_fn}
\end{figure}

For each MGC field we then computed a theoretical zero-point $ZP_{\rm
th} = a + a_{\rm am} \: \sec Z$, which is expected to be correct only
for those fields observed during the photometric night. We extracted
objects as described in Section \ref{objex} and identified duplicate
detections in regions where any two images overlapped. Using only
objects with $17 \le \bdmgc \le 21$~mag and stellaricity $>0.5$ 
(cf.\ Section \ref{class}) we computed, for each overlap region, the 
median of the magnitude differences of the double detections, $\langle
\Delta\bdmgc \rangle$, as well as an error on the median,
$\sigma_{\langle \Delta\bdmgc \rangle}$. In Fig.~\ref{dm_fn}(a) we
show $\langle \Delta\bdmgc \rangle$ for all the overlap regions using
the initial, theoretical zero-points.

A linear least-squares routine was then used to adjust the zero-points
of the non-photometric fields in order to minimize the quantity
\begin{equation}
\chi^2 = \sum \left( \frac{\langle \Delta\bdmgc \rangle}{\sigma_{\langle
\Delta\bdmgc \rangle}} \right)^2,
\end{equation}
where the sum goes over all overlap regions. The process of object
extraction, matching and zero-point adjustment was then repeated until a
stable solution was reached (four iterations).

Since the zero-points of the photometric fields are held fixed at
their theoretical values, the above procedure assumes that the
observing conditions were perfectly stable throughout the photometric
night. One can derive an estimate of the real-life error on the
photometric zero-points, $\sigma_{ZP}$, by comparing the scatter
of $\langle \Delta\bdmgc \rangle$ with $\sigma_{\langle \Delta\bdmgc
\rangle}$ for those overlap regions that only involve photometric
fields. We found $\sigma_{ZP} = 0.005$. Thus we modified the above
calibration procedure by including the zero-points of the photometric
fields in the parameters to be fitted and adding the additional
constraints that they must lie 'close' to their theoretical values. In
other words we now minimize the quantity
\begin{equation}
\chi^2 = \sum_{\rm all} \left( \frac{\langle \Delta\bdmgc
\rangle}{\sigma_{\langle \Delta\bdmgc \rangle}} \right)^2 + \sum_{\rm
phot} \left( \frac{ZP - ZP_{\rm th}}{\sigma_{ZP}} \right)^2,
\end{equation}
where the first sum again runs over all overlap regions and the second
sum runs over all photometric fields. In Fig.~\ref{dm_fn}(c) we show
$\langle \Delta\bdmgc \rangle$ for all the overlap regions using the
final zero-points. Fig.~\ref{dm_fn}(d) shows the histogram of the
individual $\Delta\bdmgc$ values. The width of this distribution
indicates an internal photometric accuracy of $0.023$~mag for objects
in the range $17 \le \bdmgc \le 21$~mag. Due to the paucity of
photometric fields beyond MGC field $74$ (cf.\ top panel of
Fig.~\ref{obsstat}) the absolute calibration in the second half of the
survey is less reliable than in the first: up to field $74$ the median
of the $\Delta\bdmgc$ distribution is $3\times10^{-4}$~mag, which is
consistent with zero, but beyond field $74$ it is $-0.004$~mag. This
is needed to reconcile the photometric fields $124$ and $139$ with the
photometric fields at field numbers $\le 74$.

Given the above calibration process the relationship between an
object's $\bdmgc$ magnitude and its Landolt $B$ magnitude
is given by:
\begin{equation} \label{coloureq}
\bdmgc = B - 0.145 (B-V),
\end{equation}
where the error on the colour term is $\pm 0.002$.

\section{Object extraction} \label{objex}
Object extraction was performed using {\sc Extractor}, which is the
STARLINK adapted\footnote{Note that the STARLINK version has
additional data handling routines, the STARLINK World Coordinate
System software and a graphical interface. In all other aspects the
code is identical to that developed and described by \citet{Bertin96},
thus henceforth we refer to {\sc SExtractor} in the text.} version of
{\sc SExtractor} developed by \citet{Bertin96}.

\subsection{Background estimation} \label{bgest}
{\sc SExtractor} initially derives a background map by first defining
a grid over the image and then passing a median filter (set to a size
of $7 \times 7$ pixel) over each pixel within the grid cell. The
local sky within the grid cell is then taken as the mode or
$\sigma$-clipped mean of the pixel distribution within the cell.
Finally, a background map is constructed via a bicubic spline
interpolation over these points. We opted for the largest possible
mesh size ($256 \times 256$ pixel) to minimize the smoothing out of
any extended low surface brightness features.

\subsection{Object detection and deblending}
After convolving the image with a filter, {\sc SExtractor} detects
objects as groups of connected pixels above the uniform surface
brightness threshold of $\mu_{\rm lim}=26$\mpass. Once an object has
been detected {\sc SExtractor} redetects the object using $30$
different detection thresholds, which are exponentially spaced between
the peak flux value of the object and the detection threshold. If at
any level the object breaks up into two or more disconnected
subcomponents, each containing at least $10$ per cent of the total
flux of the object, then the object is deblended. This
multithresholding is ideally suited for galaxy extraction because no
assumptions concerning the shape of the object are being made
\citep{Bertin96}.

\subsection{Photometry} \label{photometry}

\begin{figure*}
\psfig{file=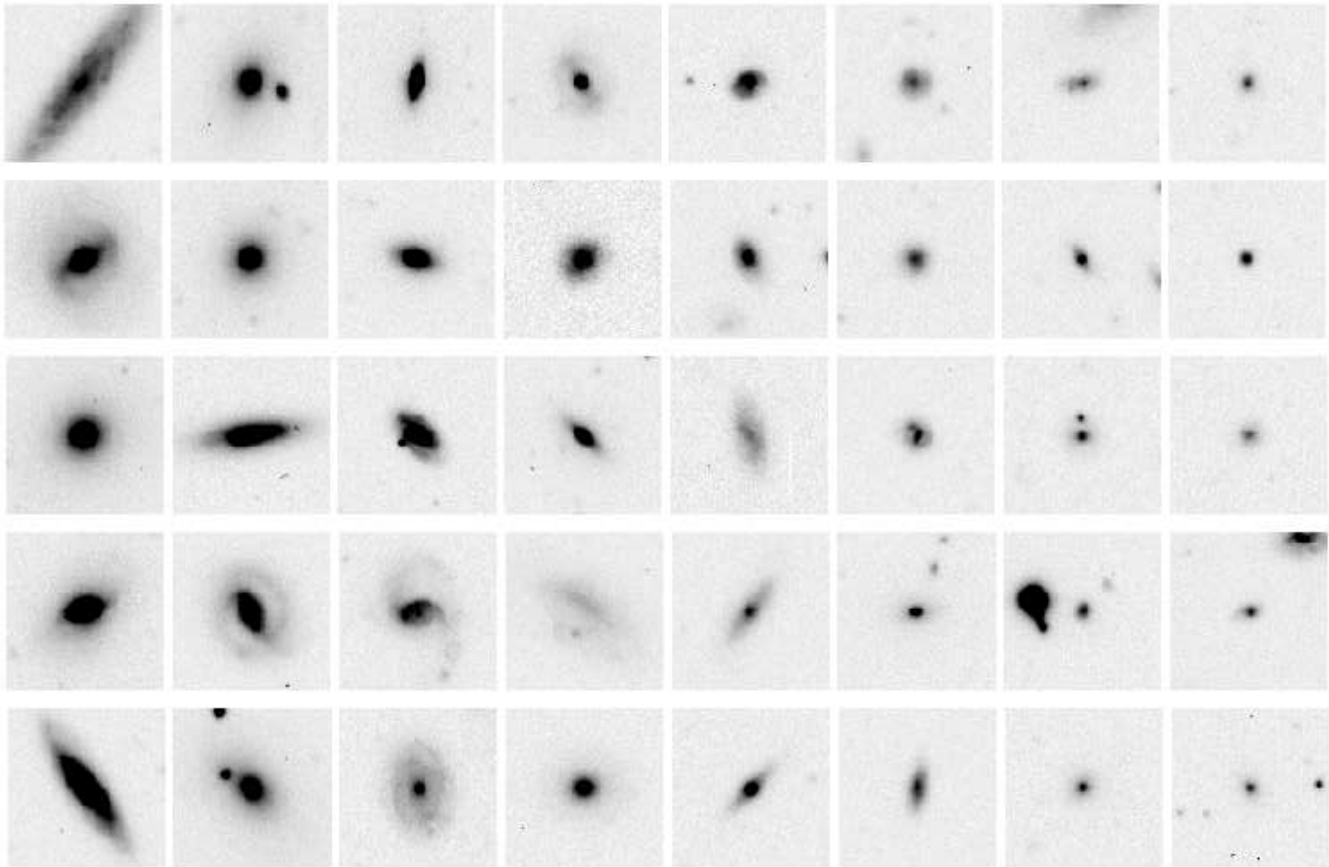,width=\textwidth,angle=90}
\caption{Random examples of the field galaxy population at $\bmgc =
16.25$~mag (left) in steps of $0.5$~mag to $\bmgc = 19.75$~mag (right)
displayed from $22$\mpass\ (black) to $3 \sigma$ below the sky
(white). The image sizes are $33 \times 33$~arcsec$^2$.}
\label{plate}
\end{figure*}

The photometry was performed using {\sc SExtractor} with a constant
analysis isophote of $26$\mpass\ to provide a uniformly processed
catalogue. Four of the various types of magnitudes provided by {\sc
SExtractor} are included in the MGC: an isophotal magnitude (ISO), a
corrected isophotal magnitude (ISOCOR), an adaptive aperture magnitude
(KRON) and a best magnitude (BEST). The ISOCOR magnitude is calculated
by correcting the ISO magnitude for the fraction of the flux outside
the limiting isophote assuming a Gaussian profile \cite[see][and
references therein for details]{Bertin96}. KRON magnitudes 
(i.e.\ elliptical apertures of 2.5 Kron radii, \citealt{Kron80}) are known to
underestimate the fluxes in perfect exponential profiles by $0.04$~mag
and in de Vaucouleur profiles by $0.1$~mag. Nevertheless, they have
been shown to be the most robust to variations in redshift,
bulge-to-disc ratio, isophotal limit and seeing \citep{Cross02b}. The
BEST magnitude is taken to be the KRON magnitude except in crowded
regions where the ISOCOR magnitude is used instead. All objects are
individually corrected for Galactic extinction using the dust
extinction maps provided by \citet*{Schlegel98} and adopting
$A_{B_{\rm KPNO}}=4.23$. From now on we refer to the BEST magnitudes
before and after extinction correction as $\bdmgc$ and $\bmgc$,
respectively.  Fig.~\ref{plate} shows a random selection of galaxies
in the range $\bmgc = 16.25$ to $19.75$~mag.

\subsection{Overlap regions}
As a result of the substantial overlap regions 
(cf.\ Fig.~\ref{outline}) the catalogue contains many duplicate
objects. These were used in previous sections to verify the astrometry
and to calibrate the photometry. We now remove duplicate detections by
imposing RA limits for each field, effectively splitting each overlap
region in half. (Note that the RA limits do not apply to objects
detected on CCD 3.) For a small number of objects, all lying very
close to an RA limit, this procedure did not remove one of the
duplicate detections, because the two detections happened to lie on
either side of the limit. These cases were fixed by hand.

\subsection{Classification and cleaning} \label{class}
\begin{figure}
\psfig{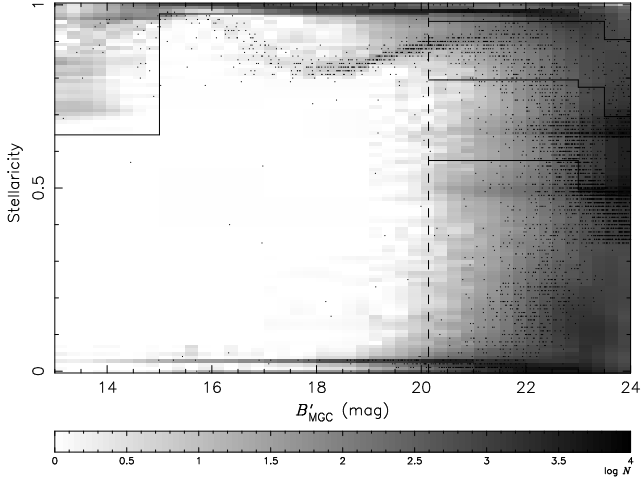}
\caption{Number of all detected objects as a function of stellaricity
and $\bdmgc$. Dots indicate the positions of objects from the field
with the very worst seeing. The vertical dashed line marks the
approximate division between MGC-BRIGHT and MGC-FAINT (which was
defined in $\bmgc$, not $\bdmgc$). In MGC-BRIGHT, the solid line
delineates the division between objects assumed to be stars and those
classified by eye. In MGC-FAINT the three solid lines show the
star--galaxy separation adopted for the fields with the best, median
and worst seeing (see the text for details), where the uppermost line
corresponds to the best seeing.}
\label{stell}
\end{figure}

At this stage the catalogue contains a total of $1\;070\;374$ objects to
$\bmgc = 24$~mag. The catalogue comprises galaxies, stars and various
unwanted objects and artefacts such as satellite trails, CCD defects,
cosmic rays, diffraction spikes, asteroids and spurious noise
detections. As a starting point for classification we used the
stellaricity parameter provided by {\sc SExtractor}, which is produced
for each object by an artificial neural network (ANN) that has been
extensively trained to differentiate between stars and galaxies
\citep{Bertin96}. The input of the ANN consists of nine object
parameters (eight isophotal areas and the peak intensity) and the
seeing. The output consists of a single number, called stellaricity,
which takes a value of $1$ for stars, $0$ for galaxies and
intermediate values for more dubious objects. Fig.~\ref{stell} shows
the number of objects as a function of stellaricity and $\bdmgc$. At
$\bdmgc \la 20$~mag the stellaricity distribution is clearly bimodal
with almost all values at the extremes and so star--galaxy separation
is trivial. At fainter magnitudes the star--galaxy separation requires
more effort. Hence we now define two catalogues: MGC-BRIGHT, which
contains all objects with $\bmgc < 20$~mag, and MGC-FAINT which
contains the rest.

\subsubsection{MGC-BRIGHT: $\bmgc < 20$}
For MGC-BRIGHT we adopted the following classification strategy.
Brighter than $\bdmgc = 15$~mag we classified all objects with
stellaricity $> 0.65$ as stars. This low stellaricity cut was used
because the objects in the upper left-hand corner of Fig.~\ref{stell}
are, in fact, flooded stars. For the rest of MGC-BRIGHT we classified
all objects with stellaricity $\ge 0.98$ as stars. This is a `natural'
value to adopt because the stellaricity distribution rises sharply
from $0.97$ to $0.98$. All objects so far classified as non-stellar
were then inspected visually and classified into one of the following
categories: galaxy, star, asteroid, satellite trail, cosmic ray, CCD
defect, diffraction spike or spurious noise detection. Incorrectly
deblended galaxies were then repaired by hand, but their original
catalogue entries were retained and classified as obsolete. Asteroids
were verified using images from the SuperCOSMOS Sky Survey
\citep[SSS,][]{Hambly01}. We also inspected all objects with FWHM,
semimajor or semiminor axis less than the seeing (these turned out
to be primarily cosmic rays and CCD defects). During this process
galaxies were also assigned one of three quality classes, $Q$,
depending on the level of: (i) contamination by CCD defects, satellite
trails, cosmic rays and diffraction spikes; (ii) blending with a
similarly bright object; (iii) missing light due to a CCD edge and
(iv) failed background estimation due to nearby bright objects. The
breakdown of the final MGC-BRIGHT catalogue is shown in
Table~\ref{brighttab} and is a good indication of the level of
contamination in purely automated galaxy catalogues.

\begin{table}
\caption{Breakdown of MGC-BRIGHT ($\bmgc < 20$~mag).}
\label{brighttab}
\begin{tabular}{lrr}
\hline
Description & Number & After cleaning \\
\hline 
Galaxies & 11866 & 9913\\
$\;\;\;(Q^a = 1,2,3$ & 11266, 449, 151 & 9775, 138, 0)\\
Stars & 51284 & 42365\\ 
Asteroids & 145 & 125\\ 
Satellite trails & 162 & 0\\ 
Cosmic rays & 113 & 62\\
CCD defects & 3027 & 0\\ 
Diffraction spikes & 263 & 0\\
Noise detections & 2023 & 13\\
Obsolete & 140 & 116\\
\hline
Total & 69023 & 52594\\
\hline
\end{tabular}

\medskip
$^a$Quality class, where 1 denotes highest quality.
\end{table}

\subsubsection{Exclusion regions} \label{exregions}
Having classified MGC-BRIGHT we now have reasonably good indicators
for the positions of CCD defects, satellite trails, very bright
objects and diffraction spikes. All of these adversely affect the
measurement of parameters of nearby objects. In addition, very bright
objects ($\bdmgc < 12.5$~mag) cause a halo of spurious faint
detections. Before we continue with the star--galaxy separation in
MGC-FAINT it is important to remove as many of these adversely
affected and spurious detections as possible. Hence we now define
exclusion regions: any objects within these regions will be removed
from both MGC-BRIGHT and MGC-FAINT to produce `cleaned' versions of
these catalogues.

In particular, we define rectangular or elliptical exclusion regions
around CCD edges, the vignetted corner of CCD 3, CCD defects, CCDs 3
and 4 of field 79 (which failed to read out properly), small, unwanted
overlaps between CCDs 2 and 3 of a few neighbouring fields, satellite
trails, diffraction spikes and objects with $\bdmgc < 12.5$~mag. For
most `classes' of exclusion regions we use some simple algorithm to
define a first set of exclusion regions from the corresponding class
of detections in MGC-BRIGHT. This first set is then improved upon and
augmented by hand where necessary. Since the parameters of very bright
objects in MGC-BRIGHT are unreliable due to saturation, we have
primarily used the SSS to define the last class of exclusion regions.

Fig.~\ref{mgc036} shows the exclusion regions for field 36. The
exclusion regions reduce the total area of the survey from
$37.50$~deg$^2$ to $30.84$~deg$^2$.

Fields 14, 15 and 65 are of substandard quality because their surface
brightness detection limit is considerably brighter than $26$\mpass
(cf.\ bottom panel of Fig.~\ref{obsstat}). This results in a large
number of spurious detections at magnitudes $\ga 23$~mag in these
fields. As part of the cleaning process we have therefore removed all
objects from these fields from MGC-FAINT (but not from MGC-BRIGHT).

\subsubsection{MGC-FAINT: $\bmgc \ge 20$} \label{mgcfaint}
\begin{figure}
\psfig{file=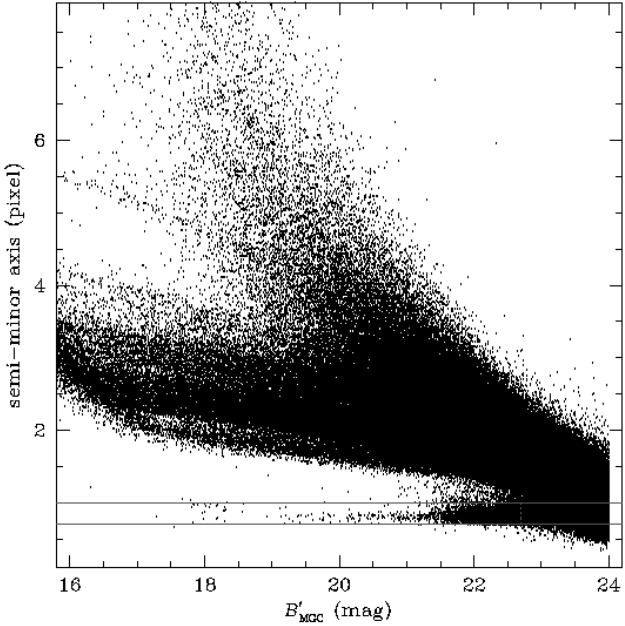,width=\columnwidth}
\caption{Flux rms along the minor axis versus $\bdmgc$ for all objects
from the cleaned versions of MGC-BRIGHT and MGC-FAINT. Almost all
cosmic rays lie in the band delineated by the two solid lines. For
$\bdmgc < 22.7$~mag the band is well separated from the general
population and hence reliable and complete cosmic ray identification
is possible using a minor axis cut alone. At $\bdmgc \ge 22.7$~mag we
apply the additional cut of FWHM $< 0.9\times \mbox{\rm seeing}$ to
identify cosmic rays but the selection is now incomplete.}
\label{min_mag}
\end{figure}

Although MGC-FAINT has been cleaned as described above, it still
contains large numbers of cosmic rays, which we now attempt to
identify. Cosmic rays are expected to be very small along at least one
axis. {\sc SExtractor} provides object shape parameters in the form of
an `rms ellipse'. In Fig.~\ref{min_mag} we plot all objects from the
cleaned versions of MGC-BRIGHT and MGC-FAINT in terms of their flux
rms along the minor axis and $\bdmgc$. We can clearly identify a `band'
of objects with $0.7 <$ minor axis $<1.0$ pixel (solid lines). From
our classification of MGC-BRIGHT and inspection of random samples at
fainter magnitudes we found that all objects within this band and
$\bdmgc < 22.7$~mag are indeed cosmic rays. Conversely, clearly almost
all cosmic rays lie in this band. However, at $\bdmgc \ga 22.7$~mag the
cosmic ray band merges with the general population of objects and a
minor axis cut alone is insufficient to select cosmic rays.

{\sc SExtractor} provides an estimate of the FWHM of the flux profile,
assuming a circular Gaussian core. The additional cut of FWHM $< 0.9
\times \mbox{\rm seeing}$ is successful in identifying large numbers of
cosmic rays at $\bdmgc \ge 22.7$~mag. The remaining unidentified cosmic
rays are those with a large angle of incidence which leave a faint
`trail' on the images. Unfortunately, given the parameters available
from {\sc SExtractor} this class of cosmic rays is genuinely
indistinguishable from real objects, and hence the identification of
cosmic rays at $\bdmgc \ge 22.7$~mag must remain incomplete. In Section
\ref{crcor} we estimate this incompleteness to be $< 20$ per cent
and correct the galaxy number counts accordingly.

All objects in the cleaned version of MGC-FAINT not identified as
cosmic rays by the above procedure are assumed to be either stars or
galaxies. As we have noted above, for $\bdmgc \ga 20$~mag star--galaxy
separation becomes increasingly difficult. However, \citet{Kummel01}
showed that the slope of the star counts remains constant to $B =
22.75$~mag, in agreement with Galaxy model predictions
\citep{Bahcall80,Bahcall86}. Hence one can produce statistical galaxy
number counts without explicitly classifying individual objects by
simply subtracting the expected stellar counts (determined by
extrapolation from the bright end) from the measured total counts.

Nevertheless, a variety of applications do require classifications for
individual objects and so we adopt the following scheme. The idea is
to find a stellaricity value, $s_{\rm c}$, such that the objects with
stellaricity $> s_{\rm c}$ reproduce the numbers of stars expected by
extrapolating the stellar counts measured in the range $17 \le \bdmgc
< 20$~mag. However, the ability of {\sc SExtractor} to distinguish
between stars and galaxies depends both on magnitude and the
seeing. Moving to fainter magnitudes or worse seeing both have the
effect of redistributing stars from the high end of the stellaricity
distribution towards intermediate values. Therefore, we must determine
$s_{\rm c}$ as a function of $\bdmgc$ for each MGC field individually.

Since for a given field $s_{\rm c}(\bdmgc)$ should be monotonic and
since the exact placement of $s_{\rm c}$ is most crucial at the faint
end, we have adopted the following strategy: for each field we first
adjust $s_{\rm c}$ in the field's faintest three bins (determined by
the field's completeness limit, see Section \ref{completeness}) to
give star count values closest to the extrapolated bright star count
fit (in a $\chi^2$ sense), while requiring $s_{\rm c}(\bdmgc)$ to be
monotonic. In all remaining bins with $\bdmgc \ge 20$~mag we set
$s_{\rm c}$ to the highest $s_{\rm c}$ value so far obtained. The
bright star count fit is derived locally for each field from the
stellar counts in the range $17 \le \bdmgc < 20$~mag as measured from
the five fields centered on the field under consideration (to improve
the reliability of the fit). In Fig.~\ref{stell} we show the resulting
$s_{\rm c}$ curves for the fields with the best, median and worst
seeing. As expected we found that the value of $s_{\rm c}$ correlates
strongly with seeing.

The error on the galaxy number counts introduced by the uncertainty
in the exact placement of $s_{\rm c}$ is discussed in Section
\ref{error}.

\subsection{Faint completeness limits} \label{completeness}
In order to estimate the faint completeness limit for each field we
used the {\sc artdata} package of {\sc IRAF} to add $100$ equally
bright point sources at random positions (but avoiding exclusion
regions) to each CCD of each field. We then re-extracted object lists
to determine whether the simulated objects were recovered. Detections
that were judged to be due to a blend of a simulated object and a real
object of similar or greater brightness were not counted as
recovered. This process was repeated for a total of $14$ input
magnitudes in the range $22 \le \bdmgc < 26$~mag. For each field we
then fitted a three-parameter completeness function to the fraction of
recovered objects as a function of magnitude \citep[cf.][]{Kummel01}:
\begin{equation}
c(\bdmgc) = \left[ \exp\left(\frac{\bdmgc - B^\prime_{50}}{\beta}\right)
+ \frac{1}{c_{\rm b}} \right]^{-1},
\end{equation}
where $c_{\rm b}$ is the completeness at bright magnitudes and
$B^\prime_{50}$ is the magnitude at which the completeness reaches
$c(B^\prime_{50}) = c_{\rm b} (c_{\rm b} + 1)^{-1} \approx 0.5$. In
general $c_{\rm b} \neq 1$ because some small fraction of objects will
always be covered up by brighter objects and thus go undetected.

We find that $c_{\rm b}$ is weakly correlated with seeing, varying
from 0.995 to 0.965 with a mean value of $\overline{c_{\rm b}} =
0.983$. $B^\prime_{50}$ is quite strongly correlated with seeing while
$\beta$ correlates with the sky noise ($\overline{\beta} = 0.22$).

For each field we now define a magnitude limit by $c(B^\prime_{\rm
lim}) = f c_{\rm b}$, where we take $f = 0.97$ so that the mean
completeness at the magnitude limit is $\overline{c(B^\prime_{\rm
lim})} = f \overline{c_{\rm b}} > 0.95$. Finally, for a given field we
define a dust-corrected magnitude limit, $B_{\rm lim}$, as the
brightest $\bmgc$ of all objects with $\bdmgc \ge B^\prime_{\rm lim}$
in that field. In the following we will only use objects with $\bmgc <
B_{\rm lim}$ and we will not attempt to use fainter objects in
combination with incompleteness corrections. Hence, when constructing
the number counts we will only use a given field for a given magnitude
bin if the field covers the entire bin (of size $\Delta \bmgc =
0.5$~mag). Thus we are effectively rounding the magnitude limit of the
field downwards to the nearest multiple of $0.5$. This leaves $124$
fields in the $\bmgc = 23.25$~mag bin and $14$ fields in the faintest
bin at $\bmgc = 23.75$~mag.

\subsection{Object distribution in RA}
\begin{figure}
\psfig{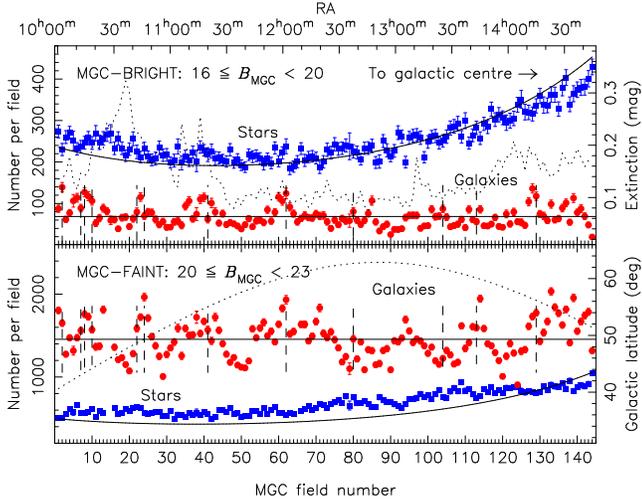}
\caption{The distribution of galaxies and stars in MGC-BRIGHT (top
panel) and MGC-FAINT (bottom panel) as a function of MGC field number
or RA (indicated along top). The dashed vertical lines indicate the
positions of $z < 1$ galaxy clusters found in the literature. The
galaxies show a uniform distribution whereas the stars show a clear
rise towards the Galactic Centre. The solid horizontal lines indicate
the mean number of bright and faint galaxies per field. The curved
solid lines represent the star counts predicted by the standard Galaxy
model of \citet{Bahcall80} and \citet{Bahcall86}. All numbers have
been scaled to a constant area of $0.215$~deg$^2$ per field. The
dotted lines and right axes indicate how the extinction and galactic
latitude vary across the survey. }
\label{radist}
\end{figure}

Fig.~\ref{radist} shows how the MGC-BRIGHT (upper panel) and MGC-FAINT
(lower panel) galaxies are distributed along the MGC survey strip. The
vertical dashed lines indicate the positions of known $z < 1$ galaxy
clusters. It is encouraging to note that on the whole they coincide
with peaks in the galaxy numbers.

We also show how the numbers of stars vary across the survey strip and
how they compare to predictions of the standard Galaxy model of
\citet{Bahcall80} and \citet{Bahcall86}. The basic shape of the counts
and the model agree in that there is a clear rise towards the galactic
bulge. However, there appear to be significant differences. The model
systematically under-predicts the faint counts and there are also
discrepancies with the bright counts at low galactic
latitudes. Similar trends are apparent in the stellar $r^*$ counts of
\citet{Yasuda01}. We will address this issue in more detail in a
future paper \citep{Lemon03}.

\section{Selection limits} \label{sellimits}
\begin{figure}
\psfig{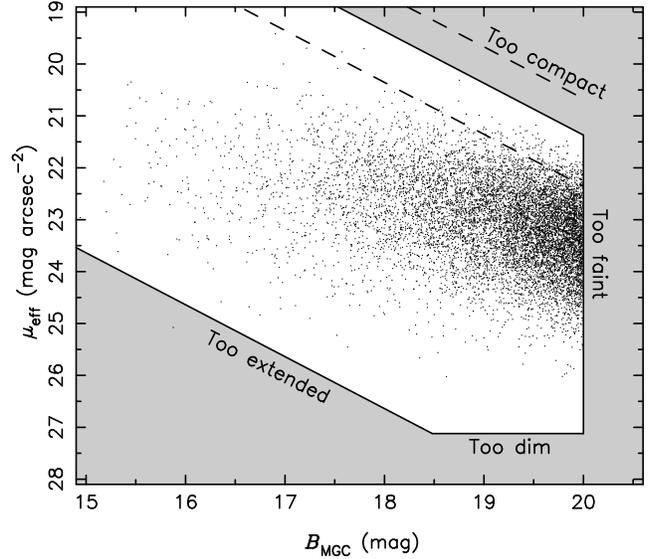}
\caption{The apparent bivariate brightness distribution of MGC-BRIGHT
galaxies. The shaded regions show the selection boundaries determined
by the median seeing (upper diagonal line), the magnitude limit
(vertical line), the surface brightness detection limit (horizontal
line) and the mesh size employed during the background estimation
(lower diagonal line, see the text for details). The two dashed lines
are the selection limits for the fields with the worst (lower) and the
best (upper) seeing.}
\label{appbbd}
\end{figure}

One of the primary aims of the MGC was to define a sample of galaxies
with well-defined selection criteria, hence the use of a constant
detection isophote. Fig.~\ref{appbbd} shows the distribution of
galaxies in the apparent magnitude -- apparent effective surface
brightness plane for MGC-BRIGHT. The apparent effective surface
brightness was simply calculated as
\begin{equation}
\mu_{\rm eff} = \bmgc + 2.5 \log (2 \pi r_{1/2}^{2}),
\end{equation}
where $r_{1/2}$ is the semimajor axis of the ellipse containing half
the total flux (in arcsec) which was measured directly from the data.

The solid lines delineating the shaded regions in Fig.~\ref{appbbd}
show the selection boundaries. The three principal selection limits
are the median seeing limit of $1.3$~arcsec, the central surface
brightness detection limit of $\mu_{\rm lim}=26$\mpass\ (equivalent to
an effective surface brightness limit of $27.12$\mpass, assuming an
exponential profile) and the imposed magnitude limit of $\bmgc =
20$~mag. Note that this cut was set by the magnitude limit at which
star--galaxy separation is reliable.

A fourth selection limit is implied by the method of background
estimation described in Section \ref{bgest}. Any objects covering a
substantial fraction of a cell of the grid used to estimate the
background will produce an erroneously high background measurement and
hence may be missed. We estimate that objects with an isophotal area
of $\sim 25$ per cent of a grid cell will be affected. We have used a
mesh size of $256 \times 256$ pixel ($85.2 \times 85.2$~arcsec).
Assuming $r_{\rm iso} \approx 2 r_{1/2}$ this represents an upper size
limit in terms of half-light radius of $64$ pixel ($21.3$~arcsec).
This line is shown as the lower diagonal selection limit in
Fig.~\ref{appbbd}.

Outside the selection boundaries galaxies cannot, theoretically, be
detected although noise may scatter a few objects across these
boundaries. Galaxies with parameters inside the boundaries should be
detectable. The galaxy population follows a well-defined distribution,
which does not reach to the high and low surface brightness selection
boundaries, demonstrating the robustness of MGC-BRIGHT with respect to
surface brightness selection effects. The distribution in surface
brightness of the observed population is far too narrow to be
explained by visibility theory \citep[see][]{Cross02} and one must
conclude that luminous low surface brightness galaxies are indeed rare
as suggested by \citet{Driver99}, \citet{Cross01} and
\citet{Blanton01}.

\section{Number counts} \label{counts}
\subsection{Errors} \label{error}
\begin{figure}
\psfig{file=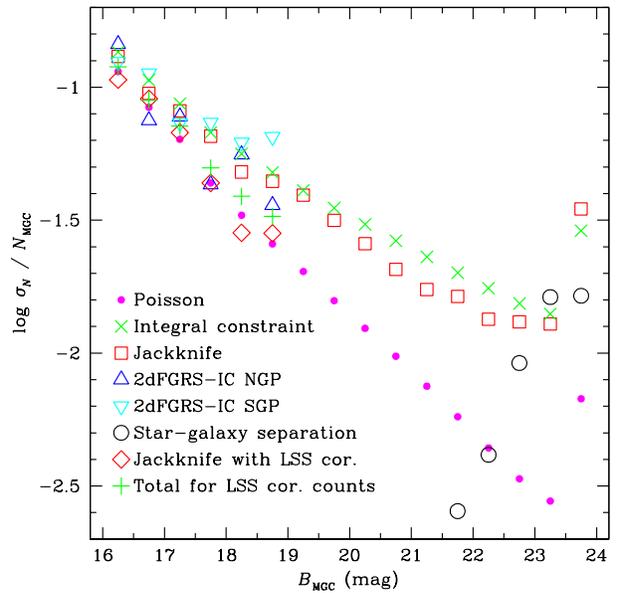,width=\columnwidth}
\caption{Various estimates of the error on the MGC galaxy number
counts, $\sigma_N$, scaled by the number counts, $N_{\rm MGC}$, as
indicated and described in the text. The last two sets refer to the
MGC-2dFGRS counts discussed in Section \ref{lsscor}. The crosses and
`+' symbols show the final errors listed in Table \ref{countstab}.}
\label{errors}
\end{figure}

Before we present the final galaxy counts we will estimate realistic
errors for the counts. In addition to Poisson noise they should also
contain a contribution from large-scale structure (LSS), which will
also induce correlations between different bins. If galaxies have an
angular correlation function $w(\theta) = A_w \theta^{1-\gamma}$ then
the error on the counts in a given bin, $\sigma_N$, is given by
\citep{Peebles80}
\begin{equation}
\sigma_N(\bmgc) = A^{-1} \sqrt{{\cal N} + {\cal N}^2 A_w C},
\end{equation}
where ${\cal N}$ is the number of galaxies in the bin, $A$ is the
survey area and $C$ is the integral constraint given by a double
integral over the survey area:
\begin{equation}
C = A^{-2} \int\!\!\!\int \theta^{1-\gamma} {\rm d} \omega_1 {\rm d} \omega_2,
\end{equation}
where we will use the SDSS-EDR results on $w(\theta)$ of
\citet{Connolly02}, who found $1-\gamma \approx -0.7$ and measured
$\log A_w$ as a function of $r^*$. We translate these measurements to
$\bmgc$ (using equation~\ref{coloureq}, the colour equations of
\citealt{Fukugita96} and the mean galaxy colours of
\citealt{Yasuda01}) and extrapolate where necessary. The resulting
error estimates are shown as crosses in Fig.~\ref{errors}. In fact, we
follow the detailed description of \citet{Yasuda01} to calculate the
full covariance matrix for the counts, which will be needed in Section
\ref{fitphi}.

We have also estimated the covariance matrix using the jackknife
method \citep{Efron82}. This was implemented by excluding each field
from the dataset in turn and measuring the counts from the remaining
data. The covariance of these $N_{\rm F}$ measurements is then
multiplied by $(N_{\rm F}-1)^2 N_{\rm F}^{-1}$, where $N_{\rm F}$ is
the number of fields used for a given pair of bins. The result agrees
well with the integral constraint estimate above and we show the
resulting errors as open squares in Fig.~\ref{errors}.

Finally, we have estimated the errors using the 2dFGRS imaging
catalogue (2dFGRS-IC). We have divided the North and South Galactic
Pole regions (NGP and SGP) into $21$ and $32$ subregions,
respectively, each of similar size and shape as the MGC survey
region. We converted the 2dFGRS-IC magnitudes to the $\bmgc$ filter
system using $\bmgc=\bj + 0.074$ (which includes a known offset of
$0.056$~mag; cf.\ equation~\ref{survcoleq}). We then measured the
counts in each independent subregion and plot the rms of these sets
separately for the NGP and SGP as triangles in Fig.~\ref{errors}.
This is only possible at $\bmgc < 19$~mag due to the magnitude limit
of the 2dFGRS-IC \citep{Colless01}.

We note that all of the above error estimates are in reasonably good
agreement and lie well above the Poisson errors (solid
dots). Henceforth we will adopt the integral constraint estimates.

In Fig.~\ref{errors} we also plot as open circles an estimate of the
uncertainty introduced by the star--galaxy separation in MGC-FAINT. We
have varied the method of determining $s_{\rm c}$ (cf.\ Section
\ref{mgcfaint}) in numerous ways but we only found a significant
change in the number counts by systematically increasing (or
decreasing) $s_{\rm c}$ in {\em all} fields. The error shown in
Fig.~\ref{errors} is the difference in the counts arising from a
global change of $s_{\rm c}$ by $0.01$. We note that such a change
produces a severe mismatch between the measured faint star counts and
those expected from extrapolating the bright counts. Hence the derived
error is probably an overestimate. In any case, it is well below the
jackknife errors in all but one bin and we therefore do not consider
it any further.

\subsection{Correcting for large-scale structure} \label{lsscor}
Since the MGC survey region is contained within the 2dFGRS-IC NGP
region the question arises as to whether the photometric accuracy and
high completeness of the former can be combined with the large area of
the latter to derive number counts less affected by LSS and hence of
higher accuracy. The idea is to view the ratio of
$N_{\mbox{\scriptsize IC-MGC}}$, the 2dFGRS-IC counts within the MGC
region, to $N_{\rm IC}$, the counts from a much larger 2dFGRS-IC
sample, as an LSS correction factor, which can be applied to the MGC
counts to give LSS-corrected counts:
\begin{equation}
N_{\rm MGC}^{\rm cor} = \frac{N_{\rm IC}}{N_{\mbox{\scriptsize
IC-MGC}}} \; N_{\rm MGC}.
\end{equation}
Equivalently, one can view $N_{\rm MGC} / N_{\mbox{\scriptsize
IC-MGC}}$ as a factor being applied to $N_{\rm IC}$, which corrects
for incompleteness, stellar contamination, bad deblending, etc.\ in
the 2dFGRS-IC. Hence we will refer to $N_{\rm MGC}^{\rm cor}$ as the
MGC-2dFGRS counts.

\begin{figure}
\psfig{file=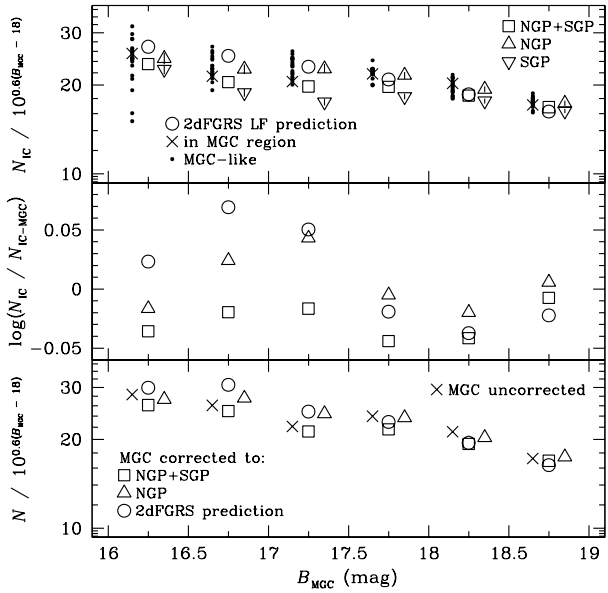,width=\columnwidth}
\caption{The top panel shows various galaxy number count estimates
derived from the 2dFGRS-IC as indicated: full sample, NGP only
($740.3$~deg$^2$), SGP only ($1094.7$~deg$^2$), in the MGC survey
region, in $21$ independent NGP subregions, each of a size and shape
similar to the MGC region. We also plot the counts predicted by the
2dFGRS luminosity function (cf.\ Section \ref{modcounts}). The error
bars on the NGP and SGP counts were calculated using the integral
constraint method. The middle panel shows the LSS correction factors
using the full, NGP only and predicted counts (same symbols as in top
panel). The bottom panel shows the uncorrected MGC counts (crosses) as
well as the MGC-2dFGRS counts using the correction factors of the
middle panel (same symbols). Several sets of counts in the top and
bottom panels have been offset by $\pm 0.1$~mag for clarity.}
\label{lsscounts}
\end{figure}

Fig.~\ref{lsscounts} illustrates the LSS correction procedure. In the
top panel we plot $N_{\mbox{\scriptsize IC-MGC}}$ as crosses along
with four sets of large-area counts for possible use as $N_{\rm IC}$
(open symbols). The middle panel shows the LSS correction factors
using the full, NGP and predicted counts (same symbols as in top
panel), and the bottom panel shows both the uncorrected MGC counts
(crosses) and the MGC-2dFGRS counts using the above correction
factors.

For this correction procedure to be valid we require only that the
properties of the 2dFGRS-IC (photometry, incompleteness, etc.) are
homogeneous over the region used to derive $N_{\rm IC}$ and the MGC
region. However, in the top panel of Fig.~\ref{lsscounts} the SGP
appears very significantly underdense in comparison to the
NGP. Indeed, \citet{Norberg02} found a $7$ per cent difference in the
numbers of NGP and SGP galaxies to $\bj = 19.2$~mag but concluded that
this was `reasonably common' in their mock galaxy catalogues derived
from $N$-body simulations. Here we take a more cautious approach and
admit the possibility of a systematic difference between the NGP and
SGP, possibly a photometric offset. Indeed, we have found an offset of
$0.056$~mag between the MGC and the 2dFGRS-IC NGP. Here and in the
previous section we have included this offset in the conversion from
$\bj$ to $\bmgc$. Applying this offset only to the NGP, but not the
SGP, makes the difference worse.

Since we require homogeneity for the LSS correction to work we choose
not to use any SGP data in $N_{\rm IC}$. On the other hand, the NGP
seems reasonably homogeneous. In the top panel of Fig.~\ref{lsscounts}
we plot as solid dots the counts from 21 independent, MGC-shaped
subregions of the NGP. We can see that the MGC region is typical and
from Fig.~\ref{errors} we have already seen that the variance among
these subregions agrees well with that expected. Hence we will use
the NGP counts as $N_{\rm IC}$.

What are the error properties of the MGC-2dFGRS counts? Defining
$f_{\rm inc} = N_{\rm MGC} / N_{\mbox{\scriptsize IC-MGC}}$ we have
\begin{equation}
\sigma_{N_{\rm MGC}^{\rm cor}}^2 = \sigma_{f_{\rm inc}}^2 N_{\rm IC}^2
+ f_{\rm inc}^2 \sigma_{N_{\rm IC}}^2 + 2 \rho f_{\rm inc} N_{\rm IC}
\sigma_{f_{\rm inc}} \sigma_{N_{\rm IC}},
\end{equation}
where $\rho$ is the correlation coefficient between $f_{\rm inc}$ and
$N_{\rm IC}$. A large value of $f_{\rm inc}$ indicates that many
galaxies in the MGC region are missed by the 2dFGRS-IC. However, this
will be true everywhere if the properties of the 2dFGRS-IC are
homogeneous. Hence $N_{\rm IC}$ will be low if $f_{\rm inc}$ is large
and vice versa, i.e.\ $\rho$ must be negative and we will err on the
side of caution by neglecting the last term. The first term can be
estimated using the jackknife method. For each `subsample' (generated
by excluding one field) we calculate $N_{\rm MGC}^{\rm cor}$. Since
$N_{\rm IC}$ does not vary from subsample to subsample the resulting
error estimate is just the first term above, which we plot as open
diamonds in Fig.~\ref{errors}. The estimate agrees very well with the
Poisson errors of $N_{\rm MGC}$ (or $N_{\rm MGC}^{\rm cor}$). Since
$f_{\rm inc} \approx 1$ we will use
\begin{equation}
\sigma_{N_{\rm MGC}^{\rm cor}}^2 = \frac{N_{\rm MGC}}{A} + 
\sigma_{N_{\rm IC}}^2,
\end{equation}
where we estimate the second term as well as the rest of the
covariance matrix using the integral constraint method. The result is
shown as `+' symbols in Fig.~\ref{errors}. We find that the errors of
the MGC-2dFGRS counts are smaller by $12$--$32$ per cent than the
errors of the counts obtained from the MGC alone.

\subsection{Correcting for asteroids} \label{astcor}
As a result of their extended morphology asteroids will be classified
as galaxies in MGC-FAINT. Recently, \citet{Ivezic01} performed
accurate measurements of the shape of the number counts of both C- and
S-type asteroids for $\bmgc \la 22.5$~mag from SDSS-CD. Here we adopt
their model, normalized to the number of visually identified asteroids
in MGC-BRIGHT, and subtract the model asteroid counts from the galaxy
counts (cf.\ Fig.~\ref{mgccounts}).

\subsection{Correcting for cosmic ray incompleteness} \label{crcor}
In Section \ref{mgcfaint} we noted that the identification of cosmic
rays is incomplete for $\bdmgc \ge 22.7$~mag. Here we will estimate the
size of the incompleteness and correct the galaxy number counts
accordingly.

\begin{figure}
\psfig{file=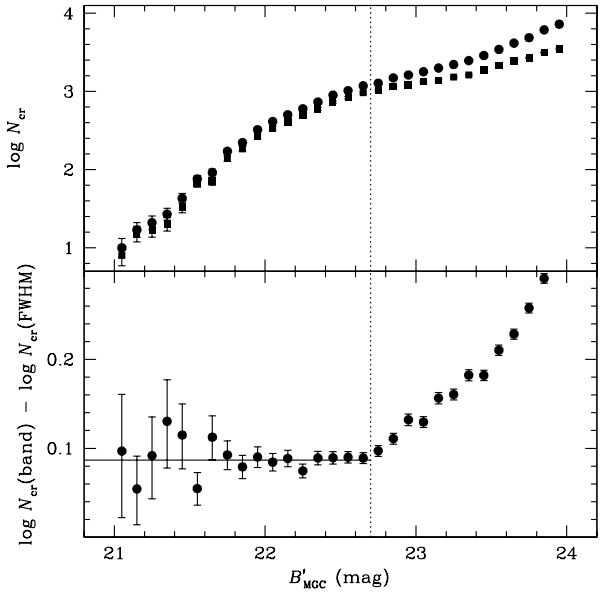,width=\columnwidth}
\caption{Upper panel: the dots show the number counts of all objects
in the cosmic ray band delineated by the two solid lines in
Fig.~\ref{min_mag}. The squares show the number counts of objects
selected by applying the additional cut FWHM $< 0.9 \times \mbox{\rm
seeing}$. Lower panel: the ratio of the above counts appears to remain
constant to $\bdmgc = 22.7$~mag (dotted line), where real objects begin
to contaminate the counts from the cosmic ray band. The incompleteness
of cosmic rays selected by the additional FWHM cut is found to be
$18.1$ per cent and is indicated by the solid horizontal line.}
\label{crcounts}
\end{figure}

In the upper panel of Fig.~\ref{crcounts} we plot as solid dots the
number counts of all objects in the cosmic ray band delineated by the
solid lines in Fig.~\ref{min_mag}. For $\bdmgc < 22.7$~mag these
represent the complete cosmic ray counts but beyond $22.7$~mag they
are `contaminated' by real objects. At $21 \le \bdmgc < 22$~mag the
counts are very steep and would exceed the galaxy counts by $\bdmgc
\approx 23$~mag if they did not flatten out. In the range $22 \le
\bdmgc < 22.5$~mag the counts indeed show some flattening. This
behaviour makes it impossible to reliably extrapolate the cosmic ray
counts to fainter magnitudes (as we do for the stars and asteroids).

We plot as solid squares the number counts of those objects that lie
in the cosmic ray band {\em and\/} have FWHM $< 0.9 \times \mbox{\rm
seeing}$, which we use as cosmic ray identification criteria at
$\bdmgc \ge 22.7$~mag. By taking the ratio of the two sets of counts
in the lower panel of Fig.~\ref{crcounts} we find that for $\bdmgc <
22.7$~mag $81.9$ per cent of all objects within the cosmic ray band
also have FWHM $< 0.9 \times \mbox{\rm seeing}$ and that this ratio
remains constant to within the errors. We now assume that this ratio
remains constant at $\bdmgc \ge 22.7$~mag and correct the cosmic ray
number counts in the range $22.5 \le \bmgc < 24$~mag by applying a
factor of $1.22$. The difference between the corrected and uncorrected
counts is then subtracted from the galaxy counts.

\subsection{Number counts} \label{ncounts}
\begin{figure}
\psfig{file=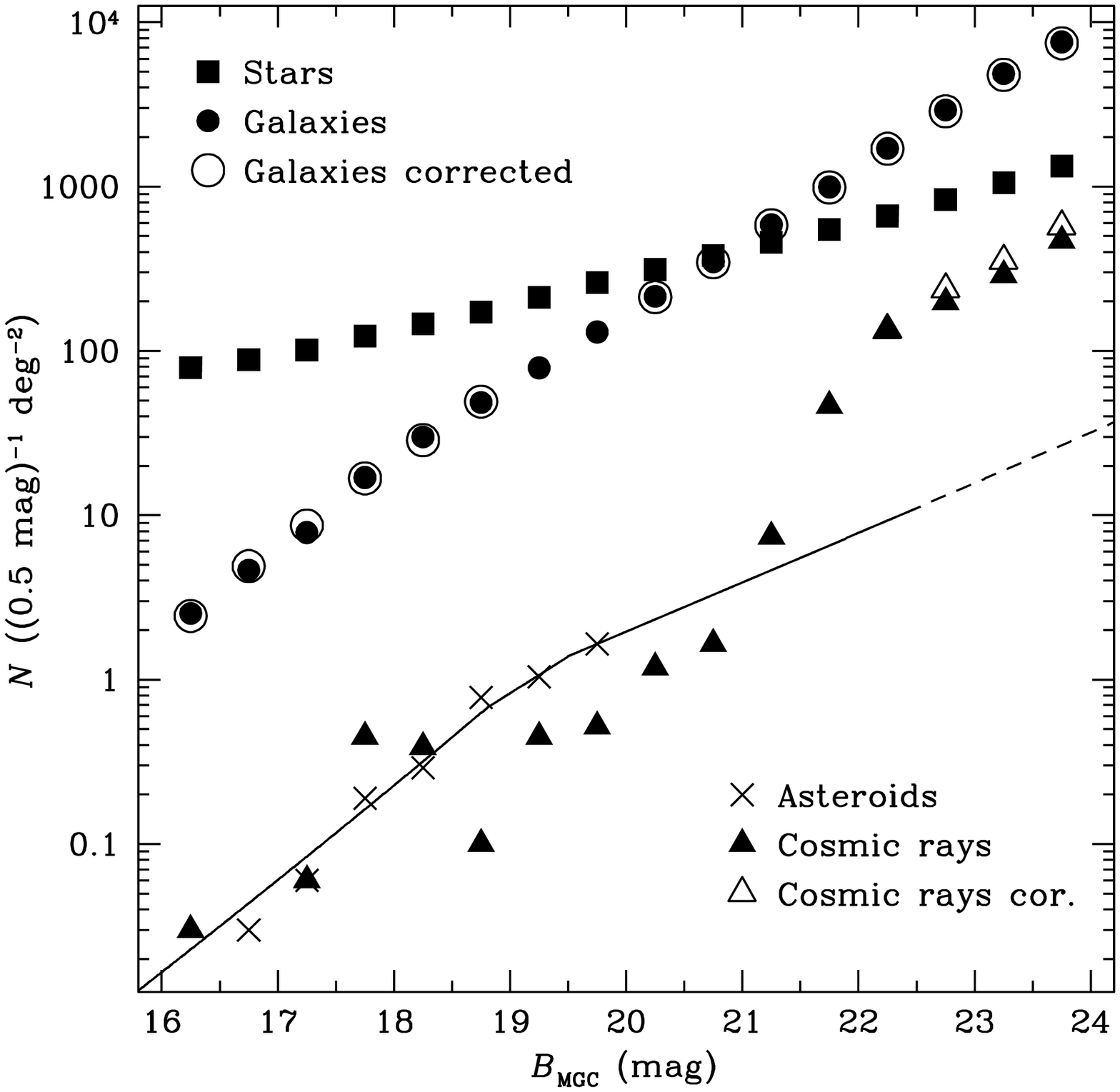,width=\columnwidth}
\caption{MGC number counts as indicated. The open circles refer to the
galaxy counts corrected for LSS ($\bmgc < 19$~mag), asteroids
($\bmgc \ge 20$~mag) and cosmic ray incompleteness ($\bdmgc \ge
22.7$~mag). The open triangles show the cosmic ray counts corrected
for incompleteness at $\bdmgc \ge 22.7$~mag. The solid line shows the
adopted \citet{Ivezic01} model for the asteroid counts, which we
extrapolate beyond the faint limit probed by these authors
(dashed line).}
\label{mgccounts}
\end{figure}

We present the MGC number counts in Fig.~\ref{mgccounts} and list the
galaxy counts in Table~\ref{countstab}. Note that the correction for
LSS (which affects the counts only in the range $16 \le \bmgc <
19$~mag) is relatively large at $\bmgc = 17.25$~mag (17 per cent)
but $\la 4$ per cent in all other bins (cf.\ middle panel of
Fig.~\ref{errors}). At $\la 1.1$ and $\sim 1.4$ per cent the asteroid
and cosmic ray incompleteness corrections (which affect the counts
only at $\bmgc \ge 20$ and $\bdmgc \ge 22.7$~mag, respectively) are
smaller still.

\begin{table}
\caption{MGC galaxy number counts.}
\label{countstab}
\tabcolsep1.5mm
\begin{tabular}{lrrd{2}d{2}d{2}d{2}}
\hline
$\bmgc$ & \multicolumn{1}{c}{${\cal N}^a$} & \multicolumn{1}{c}{Area} & 
\multicolumn{1}{c}{$N_{\rm MGC}^b$} &
\multicolumn{1}{c}{$\sigma_N^c$} &
\multicolumn{1}{c}{${N_{\rm MGC}^{\rm cor}}^d$} &
\multicolumn{1}{c}{$\sigma_{N^{\rm cor}}$}\\
(mag) & & \multicolumn{1}{c}{(deg$^2$)} & & & &\\
\hline
16.25 &     78 & 30.84 &    2.53 &   0.34 &    2.44 &   0.30 \\
16.75 &    143 & 30.84 &    4.64 &   0.49 &    4.90 &   0.41 \\
17.25 &    242 & 30.84 &    7.85 &   0.68 &    8.67 &   0.56 \\
17.75 &    523 & 30.84 &   17.0  &   1.1  &   16.77 &   0.84 \\
18.25 &    925 & 30.84 &   30.0  &   1.7  &   28.7  &   1.2  \\
18.75 &   1497 & 30.84 &   48.5  &   2.3  &   49.2  &   1.6  \\
19.25 &   2433 & 30.84 &   78.9  &   3.2  &   78.9  &   3.2  \\
19.75 &   4016 & 30.84 &  130.2  &   4.6  &  130.2  &   4.6  \\
20.25 &   6495 & 30.25 &  214.7  &   6.5  &  212.4  &   6.5  \\
20.75 &  10568 & 30.25 &  349.4  &   9.2  &  346.1  &   9.2  \\
21.25 &  17727 & 30.25 &  586    &  13    &  581    &  13    \\
21.75 &  30074 & 30.25 &  994    &  20    &  988    &  20    \\
22.25 &  51663 & 30.25 & 1708    &  30    & 1698    &  30    \\
22.75 &  88353 & 30.25 & 2921    &  45    & 2869    &  45    \\
23.25 & 129936 & 26.71 & 4865    &  68    & 4782    &  68    \\
23.75 &  22068 &  2.91 & 7581    & 219    & 7451    & 219    \\
\hline
\end{tabular}

$^a$Number of galaxies in the MGC in units of ($0.5$~mag)$^{-1}$.\\
$^b$Number counts in units of ($0.5$~mag)$^{-1}$ deg$^{-2}$.\\
$^c$Integral constraint estimate of Section \ref{error}.\\
$^d$Number counts corrected for LSS ($16 \le \bmgc < 19$~mag),
asteroids ($\bmgc \ge 20$~mag) and cosmic ray incompleteness ($\bdmgc
\ge 22.7$~mag).
\end{table}

\begin{figure}
\centerline{\psfig{file=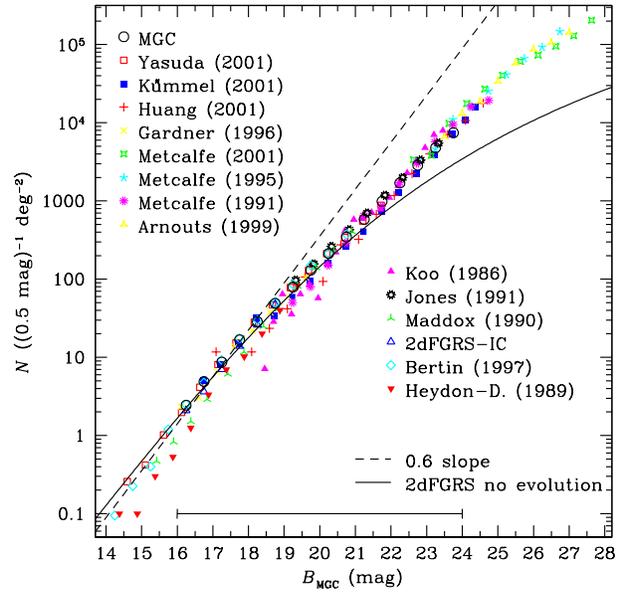,width=\columnwidth}}
\caption{The galaxy number counts derived from the MGC as compared
with the number counts of various other authors. The various counts
were converted onto the $\bmgc$ system assuming $\overline{(B-V)} =
0.94$ \citep{Norberg02}. The 2dFGRS no-evolution model counts are
calculated from the luminosity function parameters given in
Table~\ref{lumftab} (cf.\ Section \ref{modcounts}). Also shown is the
0.6 `Euclidean' slope.}
\label{allcounts}
\end{figure}

\begin{figure}
\psfig{file=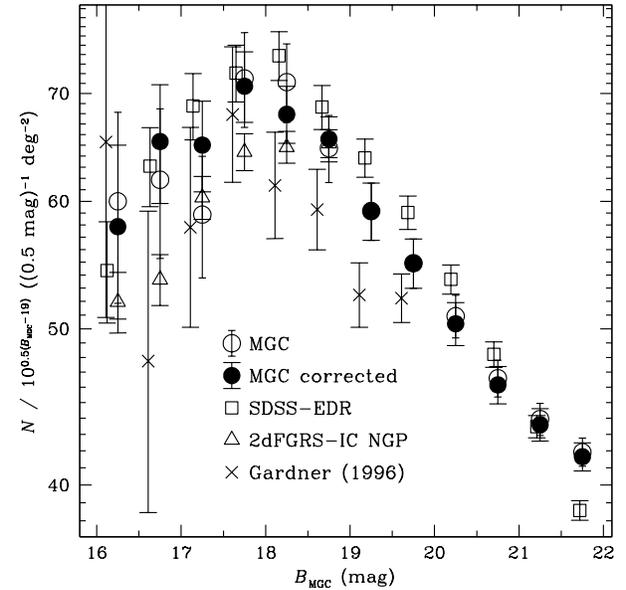,width=\columnwidth}
\caption{A close-up comparison over the important normalization range
$16 \le \bmgc < 20$~mag of the MGC number counts with three other
datasets as indicated. The MGC error bars are the integral constraint
estimates (cf.\ Sections \ref{error}, \ref{lsscor} and Table
\ref{countstab}). The SDSS-EDR and 2dFGRS-IC errors are also integral
constraint estimates, whereas the \citet{Gardner96} errors are those
published.}
\label{detailcounts}
\end{figure}

Fig.~\ref{allcounts} compares the MGC galaxy number counts to those of
previous surveys. In the range $16 \le \bmgc < 24$~mag our counts lie
among the montage of previous publications and provide a fully
consistent, uniform, well-selected and complete sample spanning eight
magnitudes. The data thus represent a significant connection between
the local photographic surveys ($B \la 20$~mag) and the deep pencil
beam CCD surveys ($B \ga 22$~mag). At bright magnitudes our counts are
consistently higher than the original APM counts \citep{Maddox90b} and
suggest that the steep rise of the APM counts is an artefact. We note
that the 2dFGRS-IC is a substantially revised version of the APM
catalogue and shows a less pronounced rise. Although some local effect
is still evident (cf.\ Fig.~\ref{detailcounts}) there is no more need
for strong local evolution of the luminous galaxy population as
originally put forward by \citet{Maddox90b}.

Fig.~\ref{detailcounts} shows a close-up comparison of the MGC-BRIGHT,
SDSS-EDR, 2dFGRS-IC NGP and \citet{Gardner96} counts, where we have
normalized the counts by an arbitrary linear model in order to make
any differences more easily discernible.

We have used the SDSS-EDR $g^*$ counts of the Northern stripe of
\citet{Yasuda01} and converted to $\bmgc$ using
equation~\eref{coloureq}, the colour equation of \citet{Fukugita96}
and the galaxy colours as a function magnitude of
\citet{Yasuda01}. Alternatively, we could have used
\citeauthor{Yasuda01}'s $B$ counts which were derived using individual
galaxy colours rather than a mean colour for a given
magnitude. However, the colour equation used by \citeauthor{Yasuda01}
is inconsistent with that of \citet{Fukugita96}. We prefer the latter
over the former because it was used in a direct comparison of SDSS-EDR
and MGC magnitudes, which showed essentially no overall zero-point
offset between the two surveys \citep{Cross03}. It was also used in
\citeauthor{Norberg02}'s comparison of the 2dFGRS and SDSS-EDR
photometry, which also showed only a small difference of $\Delta
m($2dFGRS$ - $SDSS-EDR$) = 0.058$~mag.

Fig.~\ref{detailcounts} shows that the MGC and SDSS-EDR counts are in
reasonably good agreement. In particular, both show a characteristic
change of slope near $\bmgc = 18$~mag. However, the SDSS-EDR counts
are higher than the MGC counts in the range $18 \le \bmgc < 21$~mag
by $4$--$7$ per cent. Given that \citet{Cross03} found no offset
between the MGC and SDSS-EDR photometries this difference cannot be
explained photometrically. However, the error bars in
Fig.~\ref{detailcounts}, which include an LSS component, indicate that
the difference is not significant and may well be caused by LSS.

Nevertheless, we point out that the definition of a galaxy varies
among different SDSS publications. The central tool of the SDSS-EDR
star--galaxy separation is the difference between the point spread
function (PSF) and model magnitudes, $c$, of an object. The standard
SDSS-EDR procedure is to classify an object as a galaxy if it
satisfies the condition $c > 0.145$
\citep{Stoughton02}. \citeauthor{Yasuda01} use a slight variant in
that they require $c > 0.145$ in two of the three bands $g^*, r^*$ and
$i^*$. However, \citet{Blanton01} use a cut-off value of $0.242$ in
the $r^*$ band. \citet{Strauss02} even use a value of $0.3$ in the
SDSS spectroscopic target selection in addition to a surface
brightness cut and further selection based on photometric flags. They
find that at the bright magnitudes of the spectroscopic sample ($r^*
\la 17.8$~mag) only $10$ per cent of all objects with $0.15 < c < 0.3$
are actually galaxies (see also their fig.~7) and they conclude that
their sample completeness is $> 99$ per cent.

Hence contamination by misclassified objects may in part be
responsible for the observed difference between the MGC and SDSS-EDR
counts. Indeed, the object-by-object comparison of \citet[in the range
$16 \le \bmgc < 20$~mag]{Cross03} found the SDSS-EDR galaxy sample
contaminated by stars and artefacts at the $\sim 1$ per cent level.

Comparing the uncorrected MGC and 2dFGRS-IC counts we note that the
latter are generally lower by $5$--$10$ per cent at $\bmgc <
18.5$~mag, although the difference is again only marginally
significant. We note that in the transformation from $\bj$ to $\bmgc$
magnitudes we have used the \citet{Blair82} colour term of $0.28$
(cf.\ equation~\ref{survcoleq}) and a global mean galaxy colour of
$\overline{(B-V)} = 0.94$ \citep{Norberg02}. Using instead a colour
term of $0.35$, as favoured by \citet{Metcalfe95b}, or a larger mean
galaxy colour, which may be reasonable for the fainter galaxies, both
exacerbate the difference. In any case, the small photometric offset
between the MGC and 2dFGRS-IC photometries has already been taken into
account and hence the difference cannot be explained
photometrically. However, a difference of this magnitude is easily
explained by the incompleteness and stellar contamination in the
2dFGRS-IC, which were found to be $\sim 9$, and $\sim 6$ per cent,
respectively \citep{Pimbblet01,Norberg02,Cross03}.

As far as we are aware, apart from the SDSS-EDR and the MGC the
largest CCD survey from which number count data have been published is
the $8.5$~deg$^2$ survey of \citet{Gardner96}. Although generally
lower than the MGC counts by $\sim 10$ per cent the error bars
indicate that the difference may well be due to LSS.

\section{Refining the field galaxy luminosity function} \label{lumfs}
The field galaxy luminosity function (LF) has been measured by many
surveys but, as pointed out by \citet{Cross01}, their results -- and
in particular their derived normalizations -- are inconsistent with
each other (cf.\ also Fig.~\ref{lumfmod}a). Depending on which of the
many LFs one selects, one will predict markedly different galaxy
counts, even locally (see Fig.~\ref{lumfmod}b). In the following
section we use the local galaxy counts derived in Section \ref{counts}
to assess which LFs predict counts consistent with the data and to
provide stringent constraints on the LF normalization.

\subsection{Modeling the number counts} \label{modcounts}
\begin{table*}
\begin{minipage}{\textwidth}
\caption{Parameters of published luminosity functions 
(cf.\ Fig.~\ref{lumfmod}) and their revised normalizations and luminosity
densities. For each LF we also list the probability that its number
count predictions fit the data before and after the adjustment of
$\phi^*$ (eight and seven degrees of freedom, respectively).}
\label{lumftab}
{\scriptsize
\begin{tabular}{@{\extracolsep{-3mm}}lllr@{ $\pm$\hspace{4mm}}lr@{ $\pm$\hspace{4mm}}lr@{ $\pm$\hspace{4mm}}llr@{ $\pm$\hspace{4mm}}lcr@{ $\pm$\hspace{4mm}}l} \hline
LF & Filter conversion & \multicolumn{1}{c}{$\bar z$} & \multicolumn{2}{c}{$M^*_{\bmgc} - 5 \log h$} & \multicolumn{2}{c}{$\alpha$} & \multicolumn{2}{c}{$\phi^*$} & 
\multicolumn{1}{c}{$P(\ge \chi^2_{\rm min})$} & \multicolumn{2}{c}{$\phi^*_{\rm MGC}\mbox{}^a$} & 
\multicolumn{1}{c}{$P(\ge \chi^2_{\rm min})$} & \multicolumn{2}{c}{$j_{\bj}$} \\ 
& $\bmgc=$ & & \multicolumn{2}{c}{(mag)} & \multicolumn{2}{c}{} & \multicolumn{2}{c}{($10^{-2} h^3$ Mpc$^{-3}$)} & & \multicolumn{2}{c}{($10^{-2} h^3$ Mpc$^{-3}$)} & 
& \multicolumn{2}{c}{($10^8 \; h \; L_{\odot}$ Mpc$^{-3}$)} \\ 
\hline 
2dFGRS    & $b_{\rm 2dFGRS} + 0.07$ & $0.1$   & $-19.59$ & $0.008^b$ & $-1.21$ & $0.01$ & \hspace{2mm}$1.61$ & $0.06$ & $0.18$  & \hspace{1mm}$1.69$ & $0.04\pm0.03\pm0.03$ & $0.44$ & \hspace{2mm}$2.03$ & $0.05$\\
SDSS-CD   & $g^* + 0.40$            & $0.1$   & $-19.64$ & $0.04$    & $-1.26$ & $0.05$  & $2.06$ & $0.23$ & $0$                   & $1.56$ & $0.04\pm0.11\pm0.03$ & $0.40$ & $2.06$ & $0.11$\\
ESP       & $\bj + 0.13$            & $0.1$   & $-19.59$ & $0.08$    & $-1.22$ & $0.07$  & $1.65$ & $0.3$  & $0.45$                & $1.68$ & $0.04\pm0.20\pm0.03$ & $0.42$ & $2.04$ & $0.12$\\
CS        & $V_{\rm CS} + 0.77$     & $0.064$ & $-19.53$ & $0.09$    & $-1.09$ & $0.09$  & $1.87$ & $0.21$ & $0.62$                & $1.91$ & $0.04\pm0.19\pm0.06$ & $0.60$ & $1.96$ & $0.10$\\
Dur./UKST & $\bj + 0.13$            & $0.052$ & $-19.61$ & $0.10$    & $-1.04$ & $0.08$  & $1.53$ & $0.3$  & $9 \times 10^{-7}$    & $1.78$ & $0.04\pm0.23\pm0.06$ & $0.77$ & $1.90$ & $0.09$\\
Str./APM  & $\bj + 0.13$            & $0.051$ & $-19.43$ & $0.13$    & $-0.97$ & $0.15$  & $1.26$ & $0.15$ & $0$                   & $2.20$ & $0.05\pm0.37\pm0.07$ & $0.67$ & $1.90$ & $0.13$\\
SSRS2     & $b_{\rm SSRS2} - 0.16$  & $0.02$  & $-19.61$ & $0.06$    & $-1.12$ & $0.05$  & $1.24$ & $0.2$  & $0$                   & $1.73$ & $0.04\pm0.14\pm0.07$ & $0.65$ & $1.95$ & $0.09$\\
NOG       & $B_{\rm RC3} + 0.26$    & $0.01$  & $-19.80$ & $0.11$    & $-1.11$ & $0.07$  & $1.40$ & $0.5$  & $0.87$                & $1.42$ & $0.03\pm0.20\pm0.08$ & $0.82$ & $1.89$ & $0.14$\\
\hline
\end{tabular}
}
$^a$ The three quoted errors are due to: (i) statistics and LSS, (ii)
the correlated errors on $M^*$ and $\alpha$ and (iii) the uncertainty
in $k+e$-corrections.

$^b$ We have reduced \citeauthor{Norberg02}'s zero-point uncertainty
from $0.04$ to $0.005$~mag because we have found that the 2dFGRS and
MGC photometries agree to this level once the zero-point offset has
been applied. We have also excluded the error induced by
$k+e$-corrections as we treat this error separately and consistently
for all surveys in Section \ref{fitphi}.
\end{minipage}
\end{table*}

In Table~\ref{lumftab} we list the LF parameters from the 2dFGRS
\citep{Norberg02}, SDSS-CD \citep{Blanton01}, ESP \citep{Zucca97}, CS
\citep{Brown01}, Durham/UKST \citep{Ratcliffe98}, Mt Stromlo/APM
\citep{Loveday92}, SSRS2 \citep{Marzke98} and NOG \citep{Marinoni99}
surveys. All $M^*$ values have been converted to the $\bmgc$ system
using equations~\eref{survcoleq} below. These were derived from
equation~\eref{coloureq} and the colour equations of \citet{Blair82},
\citet{Fukugita96} (SDSS-CD), \citet{Brown01} (CS), \citet{Alonso94}
(SSRS2), \citet*{Kirshner78} and \citet{Peterson86} (NOG). We also
correct for a known zero-point offset in the 2dFGRS
photometry:\footnote{Although the MGC was used during the 2dFGRS
photometry recalibration procedure (100k public release) to establish
non-linearity corrections, it was never used in the determination of
the overall zero-point of the 2dFGRS. Hence a zero-point offset
between the two surveys is possible. We have performed an
object-by-object comparison of the photometry of 5996 galaxies in
common to the two surveys and found $\Delta m ($MGC$ - $2dFGRS$) =
(-0.056 \pm 0.005)$~mag. Similarly, combining
\citeauthor{Cross03}'s \citeyearpar{Cross03} $\Delta m ($MGC$ -
$SDSS-EDR$) = -0.002$~mag with \citeauthor{Norberg02}'s $\Delta m
($SDSS-EDR$ - $2dFGRS$) = -0.058$~mag gives $\Delta m ($MGC$ -
$2dFGRS$) = -0.060$~mag.}
\begin{eqnarray} \label{survcoleq}
\bmgc &=& b_{\rm 2dFGRS} - 0.056 + 0.14 (B-V) \nonumber\\
\bmgc &=& \bj + 0.14 (B-V) \nonumber\\
\bmgc &=& g^* + 0.12 + 0.30 (B-V) \nonumber\\
\bmgc &=& b_{\rm SSRS2} - 0.02 - 0.145 (B-V) \\
\bmgc &=& V_{\rm CS} - 0.072 (V-R) + 0.8553 (B-V) \nonumber\\
\bmgc &=& B_{\rm RC3} + 0.74 - 0.51 (B-V) \nonumber
\end{eqnarray}
We assume $\overline{(B-V)} = 0.94$ \citep{Norberg02} and
$\overline{(V-R)} = 0.53$ \citep{Brown01}.

\begin{figure*}
\centerline{\psfig{file=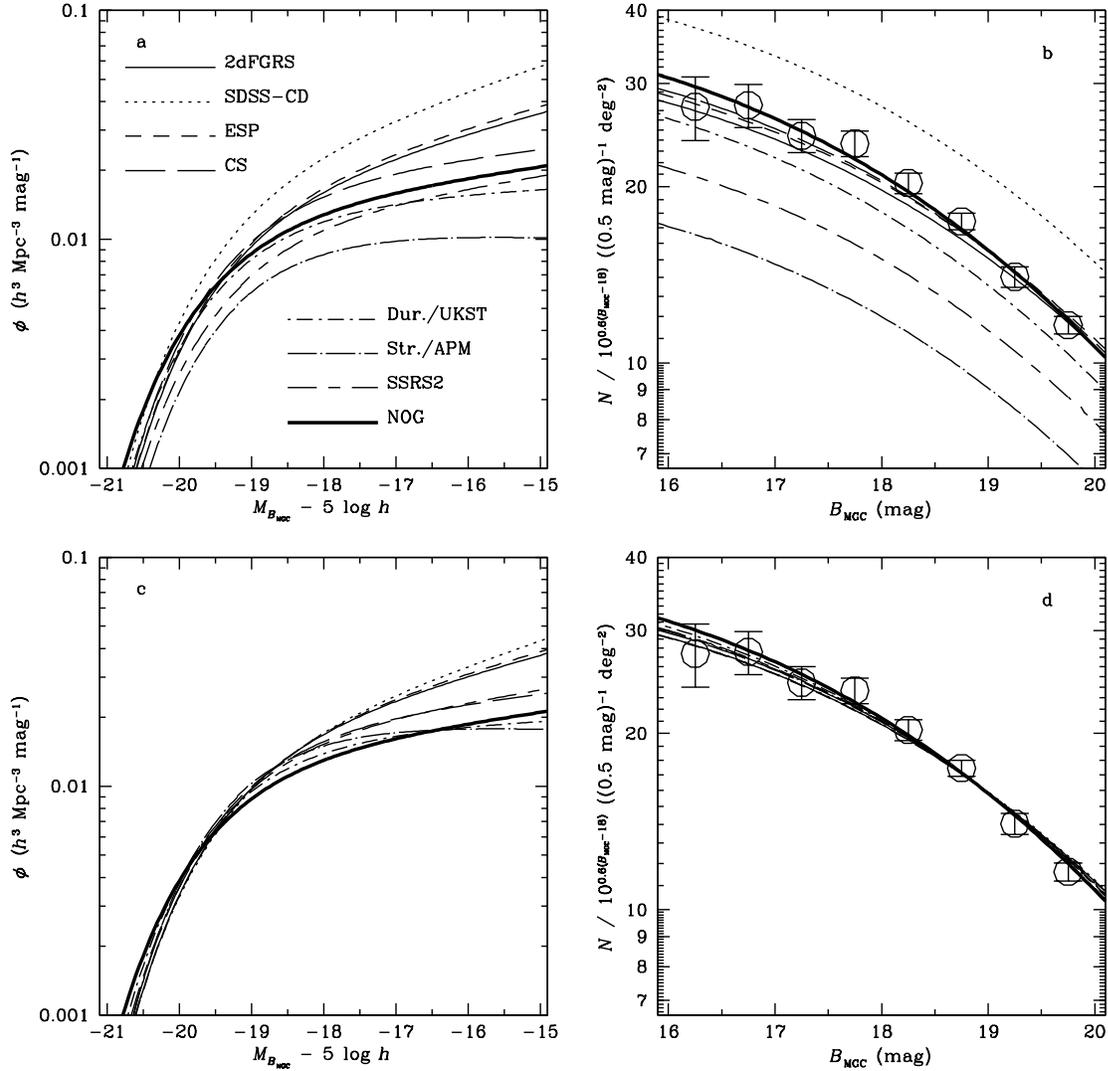,width=0.9\textwidth}}
\caption{The left-hand panels show local LFs before (upper) and after
(lower) renormalization to the corrected MGC-BRIGHT counts. The
right-hand panels show the number count predictions of these LFs and
the LSS-corrected MGC-BRIGHT counts (i.e.\ the MGC-2dFGRS counts at
$\bmgc < 19$~mag). After the recalculation of the LF normalizations,
$\phi^*$, we note that all LFs match the data equally well.}
\label{lumfmod}
\end{figure*}

Where available we have used the parameters for the currently favoured
$(\Omega_{\rm M}, \Omega_\Lambda) = (0.3, 0.7)$ cosmology (2dFGRS,
SDSS-CD and CS). The NOG parameters do not depend on the cosmological
model \citep{Marinoni98}, the SSRS2 used $(0.4, 0)$ and the rest used
$(1, 0)$. We have attempted to correct the $M^*$ values of these
latter surveys to a $(0.3, 0.7)$ cosmology using
\begin{equation}
M^*(0.3, 0.7) = M^*(\Omega_{\rm M}, \Omega_\Lambda) + 5
\log\left[\frac{r_{\rm L}(\Omega_{\rm M}, \Omega_\Lambda; \bar
z)}{r_{\rm L}(0.3, 0.7; \bar z)}\right],
\end{equation}
where $r_{\rm L}$ and $\bar z$ are the luminosity distance and median
redshift of the survey, respectively. Since all of these surveys fixed
$\phi^*$ by using one of the \citet{Davis82} estimators of the mean
galaxy density we also corrected their $\phi^*$ values using
\begin{equation}
\phi^*(0.3, 0.7) = \phi^*(\Omega_{\rm M}, \Omega_\Lambda)
\frac{\frac{{\rm d}V}{{\rm d}z}(\Omega_{\rm M}, \Omega_\Lambda; \bar
z)}{\frac{{\rm d}V}{{\rm d}z}(0.3, 0.7; \bar z)},
\end{equation}
where d$V$/d$z$ is the comoving volume element.
Since the 2dFGRS, SDSS-CD and CS have all derived their LF parameters
for several different cosmological models we can use these to test
this simple correction procedure. In all cases we find good agreement
(well within the quoted error bars) between the transformed and
measured values. The final corrected $\bmgc$ LFs are shown in
Fig.~\ref{lumfmod}(a).

To model the counts we use ($\Omega_{\rm M}, \Omega_{\Lambda}) = (0.3,
0.7)$, a $k$-correction of $2.5z$ and an $e$-correction of $2.5 \log
[(1+z)^{-0.75}]$. At $z < 0.4$ this combination matches well the
$k+e$-correction given by \citet{Norberg02} (their fig.~8) which was
derived using \citet{Bruzual93} models to match the colour--redshift
trend seen in the 2dFGRS (using SDSS-EDR colours). In principle, for
each of the surveys in Table \ref{lumftab} we should use the same
$k(+e)$-correction as was used in the derivation of the Schechter
parameters of that survey or else correct for the use of a different
$k+e$. However, in several cases different $k$-corrections were used
for different galaxy types, sometimes interpolating between types, and
hence a mean correction is not readily available.

Fig.~\ref{lumfmod}(b) shows how the counts predicted by the various
LFs compare with the MGC-BRIGHT counts (corrected for LSS). Clearly,
the data prefer some models over others. We have performed a
goodness-of-fit test for the predicted counts in the range $16 \le
\bmgc < 20$~mag (using the full covariance matrix of the counts) and
list the resulting probabilities in column 7 of
Table~\ref{lumftab}. The model counts based on the Durham/UKST, Mt
Stromlo/APM and SSRS2 LFs are all significantly too low, while the
SDSS-CD model counts are too high. The other models provide good
fits. These conclusions do not change if we compare the models to the
uncorrected counts.

From Fig.~\ref{lumfmod}(b) it is clear that the main difference
between the various predicted counts and the data lies in their
normalization, $N_{18}$, and not their shape. As discussed in Section
\ref{introduction} it is precisely this uncertainty in $N_{18}$ that
is commonly referred to as the `normalization problem'.

\subsection{The normalization problem}
For a given set of Schechter parameters what is the uncertainty in
$N_{18}$ induced by the error of each of the parameters? In the
general case there is no analytic formula relating $N_{18}$ with the
Schechter parameters. Hence the `Euclidean' case must serve as a
guideline. Ignoring binning effects we have $N_{18} \propto \phi^*
{L^*}^{3/2} \Gamma(\alpha + \frac{5}{2})$. Using the intersurvey rms
of each parameter as an indication of its uncertainty we find
\begin{equation} \label{nerror}
\frac{\partial N_{18}}{\partial \phi^*} \Delta \phi^* : \frac{\partial
N_{18}}{\partial M^*} \Delta M^* : \frac{\partial N_{18}}{\partial
\alpha} \Delta \alpha= 13 : 10 : 1
\end{equation}
and similar or worse values for the individual surveys. Hence we
identify the errors in both $\phi^*$ and $M^*$ as the main causes of
the uncertainty in $N_{18}$. What are the sources of these
uncertainties?

\subsubsection{Photometric error}
In addition to the difficulty of reliably calibrating the photometry
of large photographic surveys \citet{Cross02} showed that the derived
LF parameters depend on the limiting isophote and the photometric
method (e.g.\ isophotal, corrected, total, etc.). They concluded that
when using isophotal magnitudes for a limiting isophote of $25$\mpass\
one might expect errors in $M^*$ of up to $\pm 0.4$~mag, in $\phi^*$
of up to $\pm 10$ per cent and in $\alpha$ of up to $\pm 0.01$. Hence
photometric uncertainties contribute substantially to the
normalization problem.

\subsubsection{Incompleteness}
Although the underestimation of the magnitudes due to the limiting
isophote will cause some galaxies to fall below the limiting magnitude
of a survey, it also causes the derived volume over which such
galaxies can be seen to be {\em underestimated}. In addition,
Fig.~\ref{appbbd} (together with visibility theory) shows that at
$\bmgc \la 19$~mag apparently no significant galaxy population below
$25$\mpass\ exists. In other words, at these surface brightness limits
it is typically a case of missing light from the outer isophotes
rather than missing galaxies: \citet{Cross03} found only $\sim 0.5$
per cent of galaxies missing from the 2dFGRS due to their low surface
brightness. However, blended and unresolved objects take the total
incompleteness of the 2dFGRS to $\sim 9$ per cent, which may be
indicative of photographic surveys in general.

\subsubsection{Large-scale structure}
LSS will clearly affect all but the very largest surveys.
\citet{Norberg02} found that the $173$~deg$^2$ overlap region between
the SDSS-EDR and the 2dFGRS was overdense by $5$ per cent relative to
the full 2dFGRS. Similarly they found a $7$ per cent difference (to
$\bj = 19.2$) between their $740$~deg$^2$ NGP and $1095$~deg$^2$ SGP
regions. Hence it is clear that much shallower surveys like the SSRS2
($b_{\rm lim} = 15.5$~mag) cannot accurately measure $\phi^*$ even if
they have a very large angular size. On the other hand, due to the
filamentary structure of the Universe even deep surveys are
susceptible to LSS if they extend only over a small solid angle.

\subsubsection{$k(+e)$-corrections}
Finally, we remark that each of the surveys listed in Table
\ref{lumftab} uses different $k$-corrections, and only the 2dFGRS use
$e$-corrections. Indeed, evolutionary effects, which were ignored by
\citet{Blanton01}, are the reason why the SDSS-CD normalization is so
high. When evolution is included (or when normalizing to the SDSS-EDR
counts instead of using the method of \citealt{Davis82}) the SDSS-CD
normalization agrees much better with the 2dFGRS results
\citep{Yasuda01,Norberg02,Blanton02}.

\subsection{Determining $\phi^*$} \label{fitphi}
\begin{figure}
\psfig{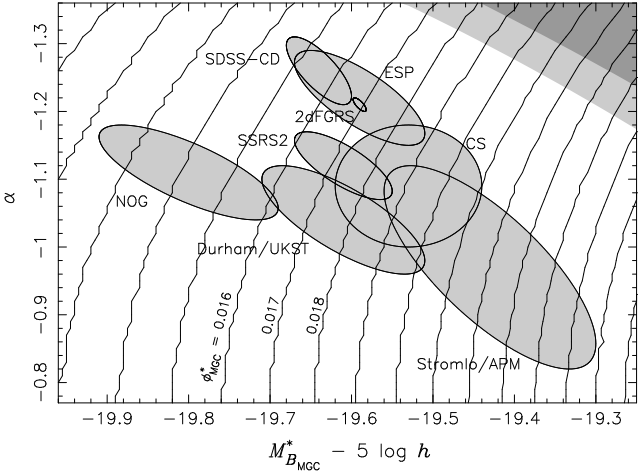}
\caption{The contours show the best-fitting value of the LF
normalization, $\phi^*_{\rm MGC}$, for given values of $M^*$ and
$\alpha$. These were derived by $\chi^2$-minimization over the range
$16 \le \bmgc < 20$~mag (cf.\ Fig.~\ref{lumfmod}, using the full
covariance matrix). The values of the contours increase from
$0.011$~$h^3$~Mpc$^{-3}$ on the left in steps of $0.001$ to $0.027$ on
the right as indicated. The error ellipses of the LFs of
Table~\ref{lumftab} in the $M^*$-$\alpha$ plane are also indicated
(assuming a correlation coefficient of $0.75$). The corresponding
best-fit $\phi^*_{\rm MGC}$ values are listed in
Table~\ref{lumftab}. The light and dark shading in the upper right
corner marks those regions in the $M^*$--$\alpha$ plane where the
shape of the predicted counts disagrees with the data at the $90$ and
$95$ per cent levels, respectively.}
\label{phistar}
\end{figure}

Although both $M^*$ and $\phi^*$ appear to contribute equally to the
uncertainty in $N_{18}$ it is customary (and sensible) to fix $M^*$
and use $N_{18}$ to constrain $\phi^*$ alone instead of a combination
like $\phi^* {L^*}^{3/2}$. Over the range $16 \le \bmgc < 20$~mag the
MGC arguably provides the most reliable number count data in existence
and hence it provides very reliable constraints on $\phi^*$. We now
determine $\phi^*_{\rm MGC}$ as a function of $M^*$ and $\alpha$ by
fitting the model counts for a given combination of $M^*$ and $\alpha$
to the corrected MGC-BRIGHT counts over the range $16 \le \bmgc <
20$~mag (i.e.\ the MGC-2dFGRS counts at $\bmgc < 19$~mag), where
$\phi^*$ is the free parameter. We use the full covariance matrix for
the fits. Fig.~\ref{phistar} shows the result as a contour plot in the
$M^*$--$\alpha$ plane.

In particular, we derive the $\phi^*_{\rm MGC}$ values appropriate for
the LFs listed in Table~\ref{lumftab}. The revised values and their
associated probabilities are listed in columns 7 and 8 of that
table. Figs.~\ref{lumfmod}(c) and (d) show the resulting LFs and their
predicted counts after revision. All surveys now give comparable and
reasonable probabilities. Note, however, that the variation of
$\phi^*$ among the different surveys after renormalization is no less
than before. This is due to the fact that the counts constrain the
combination $\phi^* {L^*}^{3/2}$ and that the variation in
${L^*}^{3/2}$ is comparable to that of $\phi^*$ before
renormalization (as seen from equation \ref{nerror}).

In Table~\ref{lumftab} we also list the errors on $\phi^*_{\rm MGC}$
induced by various sources. First, we list the statistical error
derived from the $\chi^2$ fit. This error includes remaining
uncertainties due to LSS because we have used the full covariance
matrix of the counts in the fit.
Secondly, we list the errors due to the uncertainties in $M^*$ and
$\alpha$, which can be essentially read off from
Fig.~\ref{phistar}. Here we take into account that the errors on $M^*$
and $\alpha$ are generally highly correlated by assuming a correlation
coefficient of $0.75$ \citep{Blanton01}. Finally, we have estimated
the errors resulting from uncertainties in the $k+e$-correction. This
error is slightly more subtle than the previous ones. For example,
increasing $k+e$ has the effect of moving the contours in
Fig.~\ref{phistar} to the left and up but it also moves the LFs in a
similar direction, thus again decreasing the original change in
$\phi^*_{\rm MGC}$ for a given survey. \citet{Norberg02} estimated
errors on their $k+e$-correction by demanding statistical consistency
between the LFs derived from their high and low-$z$ samples. Since our
$k+e$ is essentially identical to theirs we also adopt their error of
$\pm 18$ per cent. We estimate the resulting $\phi^*_{\rm MGC}$ error
for a given survey by first adding $\pm 0.18 (k+e)(\bar z)$ to its
$M^*$ value and then re-deriving $\phi^*_{\rm MGC}$ while using the
increased (or decreased) $k+e$. As in Section \ref{modcounts} we thus
approximate the effect on $M^*$ by evaluating $\Delta(k+e)$ at the
median redshift of the survey. Comparison with
\citeauthor{Norberg02}'s $k+e$ induced error on $M^*$ indicates that
this procedure underestimates the effect on $M^*$, and hence we
overestimate the error on $\phi^*_{\rm MGC}$. We also conservatively
ignore any effect on $\alpha$, which is difficult to gauge.

\subsection{The local luminosity density}
Taking $M_{\odot_{\bj}} = +5.3$~mag and using the revised LF
normalizations we calculate the luminosity density, $j_{\bj} = \phi^*
L^*_{\bj} \Gamma(\alpha+2)$. Note that we convert $M^*_{\bmgc}$ to
$M^*_{\bj}$. The calculated values of $j_{\bj}$ are listed in the last
column of Table~\ref{lumftab}, where the errors include all the $M^*$,
$\alpha$ and $\phi^*_{\rm MGC}$ uncertainties while taking into
account the correlation between the errors of $M^*$ and $\alpha$ (with
a correlation coefficient 0.75 as above) as well as the virtually
perfect correlations between the second error component of
$\phi^*_{\rm MGC}$ and the errors of $M^*$ and $\alpha$ and between
the $k+e$ errors on $M^*$ and $\phi^*_{\rm MGC}$. We find a weighted
mean $\bj$ luminosity density of $\overline{j_{\bj}} = (1.986 \pm
0.031) \times 10^8 \; h \; L_{\odot}$~Mpc$^{-3}$. Note that the 2dFGRS
and SDSS-CD, which must be considered the most reliable LFs, give a
slightly higher value of $j_{\bj} = (2.035 \pm 0.046) \times 10^8 \; h
\; L_{\odot}$~Mpc$^{-3}$.

\subsection{Constraints on $M^*$ and $\alpha$}
In the previous section we minimized $\chi^2$ as a function of
$\phi^*$ only while treating $M^*$ and $\alpha$ as fixed
parameters. $\phi^*$ only affects the normalization of the predicted
counts but not their shape which, in principle, also contains
information. However, for the LFs listed in Table \ref{lumftab} the
fits are already statistically acceptable (cf.\ column 9
of Table~\ref{lumftab}) and hence we cannot expect to derive useful
limits on any other parameters.

Nevertheless, in Fig.~\ref{phistar} we mark as grey-shaded regions
those combinations of $M^*$ and $\alpha$ where the adjustment of
$\phi^*$ does not result in an acceptable fit (at the $90$ and $95$
per cent confidence levels), i.e.\ where the shape of the predicted
counts disagrees with the data. As expected these limits cannot
exclude any of the measured values.

\section{Conclusions} \label{conclusions}
Here we have presented a detailed description of the Millennium Galaxy
Catalogue (MGC), a deep ($\mu_{\rm lim} = 26$\mpass) wide
($37.5$~deg$^2$) survey along the equatorial strip from $9^{\rm h}
58^{\rm m}$ to $14^{\rm h} 47^{\rm m}$. We have demonstrated that the
internal photometric accuracy of the MGC is $\pm 0.023$~mag and that
the astrometric accuracy is $\pm 0.08$~arcsec in both RA and
Dec. Using {\sc SExtractor} we have derived a source catalogue
containing over 1 million objects spanning the range $16 \le \bmgc <
24$~mag. All non-stellar detections brighter than $\bmgc = 20$~mag
have been visually inspected and the objects repaired where
necessary. We have taken care to exclude objects from regions where
the photometry is likely to be erroneous, resulting in a robust and
clean estimation of the galaxy number counts over the range $16 \le
\bmgc < 24$~mag. These data finally connect the faint pencil beam CCD
surveys of the past decade to the local Universe. The selection
boundaries of the MGC are well defined and to $\bmgc = 20$~mag we are
demonstrably robust to star--galaxy separation and low- and
high-surface brightness concerns. We contest that the MGC galaxy
number counts in this range are the state of the art, superseding all
previous intermediate number count data.

We use the counts to test various estimates of the galaxy luminosity
function and find that many of them predict counts where the
normalizations are inconsistent with our observations. In
Fig.~\ref{phistar} we present the best-fitting value of $\phi^*$ as a
function of $M^*$ and $\alpha$. In Table~\ref{lumftab} we list the
appropriate $\phi^*$ values for a number of popular $B$-band
luminosity function estimates.

From these revised values we constrain the $\bj$-band luminosity
density of the local Universe for each of these luminosity
functions. We find $\overline{j_{\bj}} = (1.986\pm 0.031) \times 10^8
\; h \; L_{\odot}$~Mpc$^{-3}$. The 2dFGRS and SDSS-EDR consistently
give a slightly higher value of $j_{\bj} = (2.035 \pm 0.046) \times 10^8
\; h \; L_{\odot}$~Mpc$^{-3}$.

\section*{ACKNOWLEDGMENTS}
We wish to thank all those who participated in observing the MGC
fields. The data were obtained through the Isaac Newton Group's Wide
Field Camera Survey Programme. The Isaac Newton Telescope is operated
on the island of La Palma by the Isaac Newton Group in the Spanish
Observatorio del Roque de los Muchachos of the Instituto de
Astrof\'{i}sica de Canarias. We also thank CASU for their data
reduction and astrometric calibration.

\label{lastpage}

\end{document}